\documentclass[pdflatex,sn-mathphys]{sn-jnl}

\usepackage{amsmath,amssymb,amsfonts,subfig,xcolor,soul,color,float,bm}
\usepackage{epstopdf}
\allowdisplaybreaks

\jyear{2021}%

\theoremstyle{thmstyleone}%
\theoremstyle{thmstyletwo}%

\theoremstyle{thmstylethree}%

\begin{document}

\title[ ]{Parametric roll oscillations of a hydrodynamic Chaplygin sleigh}


\author[1]{\fnm{Kartik} \sur{Loya}}

\author*[1]{\fnm{Phanindra} \sur{Tallapragada}}\email{ptallap@clemson.edu}

\affil*[1]{\orgdiv{Mechanical Engineering Department}, \orgname{Clemson University}, 
\orgaddress{ \city{Clemson}, \postcode{29631}, \state{S.C.}, \country{U.S.A}}}


\abstract{Biomimetic underwater robots use lateral periodic oscillatory motion to propel forward, which is seen in most fishes known as body caudal fin (BCF) propulsion. The lateral oscillatory motion makes slender-bodied fish-like robots roll unstable. Unlike the case of human-engineered aquatic robots, many species of fish can stabilize their roll motion to perturbations arising from the periodic motions of propulsors.  To first understand the origin of the roll instability, the objective of this paper is to analyze the parameters affecting the roll-angle stability of an autonomous fish-like underwater swimmer. Eschewing complex models of fluid-structure interaction, we instead consider the roll motion of a nonholonomic system inspired by the Chaplygin sleigh, whose center of mass is above the ground. In past work, the dynamics of a fish-like periodic swimmer have been shown to be similar to that of a Chaplygin sleigh. The Chaplygin sleigh is propelled by periodic torque in the yaw direction.  The roll dynamics of the Chaplygin sleigh are linearized and around a nominal limit cycle solution of the planar hydrodynamic Chaplygin sleigh in the reduced velocity space. It is shown that the roll dynamics are then described as a nonhomogeneous Mathieu equation where the periodic yaw motion provides the parametric excitation. We study the added mass effects on the sleigh's linear dynamics and use the Floquet theory to investigate the roll stability due to parametric excitation. We show that fast motions of the model for swimming are frequently associated with roll instability. The paper thus sheds light on the fundamental mechanics that present trade-offs between speed, efficiency, and stability of motion of fish-like robots.}


%




\maketitle

\section{Introduction}\label{sec:intro}
In this paper, we investigate the parametric roll dynamics of a hydrodynamic nonholonomic system inspired by a Chaplygin sleigh that is completely submerged in water and moves on the bed of a body of water. The Chaplygin sleigh is a well-known planar $3$ degree of freedom nonholonomic system with a configuration manifold $SE2$, see \cite{bloch03, V.Borisov2007, Neimark1972} for a review. The hydrodynamic model of the Chaplygin sleigh considered in this paper is motivated by applications to autonomous bioinspired underwater robots and robots that can crawl on the floor of a water body. The locomotion of fish and other aquatic swimmers has many desirable characteristics such as energy efficiency, agility, and stealth \cite{lauder_afm_2015, triantafyllou_afm_2016, lauder_sms_2019, Triantafyllou_AFM_2000}, which have inspired the design of many biomimetic robots such as in \cite{Barrett1996,pettersen_ieee_ram_2016, boyer_tor_2008,   zhu2019tuna, quinn_science_2021} to name a few. Simplified models of fish-like swimming are crucial for improving the design and control of such robots. The dynamics of the Chaplygin sleigh bear a surprising similarity to fish-like swimming. The nonholonomic constraint on the Chaplygin sleigh has been shown to be qualitatively similar to the vortex shedding Kutta-Joukowski condition in inviscid flows \cite{tallapragada_ACC2015, tallapragada_kelly_jcnd2017}. Further work showed that a planar swimming hydrofoil propelled by an internal rotor \cite{pt_tmech_2016} has limit cycle dynamics similar to that of a planar Chaplygin sleigh with viscous drag and periodic yaw torque \cite{ft_nody_2018, pft_dscc_2018,  pft_nody_2019}.  This similarity between the dynamics of the Chaplygin sleigh and a fish-like swimmer has led to several control models for path tracking \cite{paley_bb_2020, ft_acc_2020} and formation control \cite{paley_acc_2020} of planar fish-like robots. 

Planar simplified models completely ignore the three-dimensional roll motion of biological swimmers and bioinspired robots. A large number of aquatic animals use some form of an undulatory motion to generate the so-called body caudal fin propulsion (BCF), which causes cyclic yawing moments and sideways recoil forces owing to the mechanics of producing forward thrust. It has been understood through experiments that this makes fish roll unstable and that this instability increases their maneuverability, \cite{bandyopadhyay_2002, webb_weihs_icb_2015}. This paper proposes and investigates a more complex model of a Chaplygin sleigh whose center of mass is not at ground height with a configuration manifold $S^1\times SE(2)$. Such a rigid body, when placed on the ground (without any surrounding water), has a tendency to ``fall down'' since the upward position of the center of mass is an unstable equilibrium. When placed in water, the sleigh may be statically stable due to buoyancy depending on the height of the center of buoyancy and center of gravity. We further assume in our model that the sleigh is subject to a periodic torque in the body yaw direction that generates a propulsive thrust. The interaction of hydrodynamic effects like added mass and buoyancy with the nonholonomic constraint and periodic actuation leads to complex parametric oscillations. 

The main contribution of this paper is to show that the equation governing the roll motion of the hydrodynamic Chaplygin sleigh can be approximated as a nonhomogeneous forced Mathieu equation, with the forcing and the nonhomogenous terms being defined by the limit cycle solutions in the reduced velocity space of the planar nonholonomic sleigh. We have shown using Floquet theory that the hydrodynamic model, which incorporates only the buoyancy force as a hydrodynamic effect, has similar regions of instability in a nondimensionalized parameter space as the Mathieu oscillator. When the hydrodynamic model is improved by incorporating the added mass tensor, the existence of the resonance regions in the parameters space may disappear. The added mass tensor also has the curious effect of inducing a negative damping, leading to unbounded oscillations for slender bodies like a prolate ellipsoid. Only small regions in the parameter space for such bodies lead to bounded oscillations.  The proposed model and findings in the paper shed light on the interplay between body morphology, gaits, and control in fish locomotion and the bioinspired robots \cite{webb_weihs_cjz_1994, bandyopadhyay_2002, Colgate2004}.  The frequency and amplitude of the periodic yaw motion of the Chaplygin sleigh are related to the speed of the sleigh, and thus, the findings in the paper also shed light on the often competing relationship between the speed of swimming and roll stability.
 

Two other areas of research are related to the problem investigated in this paper, and we set the novelty of the current work within their context. The first is the classic problem of parametric roll oscillations of a surface vessel due to periodic heave oscillation or pitch oscillations \cite{paulling_rosenberg_1959, newman_1979,  nayfeh_jsr_1988, neves_oe_2006}. A simplified model for the vertical dynamics of a surface vessel is that of an inverted pendulum with possible base motion. A surface vessel subjected to waves has a periodic vertical motion (heave), inducing periodic changes to the height of the center of the buoyancy and thus the effective natural frequency. The resulting parametric oscillations can be described by a Mathieu equation. The origin of the roll oscillations investigated in this paper is distinct from these and is coupled to the nonholonomic constraint and the yaw oscillations of the swimmer. A second related class of problems comes from the dynamics of a unicycle \cite{zbm_cdc_1999, naveh_dc_1999, deluca_2000, zbm_scl_2002}. The unicycle is roll stable at high speeds due to the spin angular momentum about the pitch axis \cite{zbm_cdc_1999, zbm_scl_2002}, which is not present for the Chaplygin sleigh, and only needs to be stabilized at low speeds. This is achieved by changing the inertia tensor through an additional internal degree of freedom \cite{zbm_cdc_1999,zbm_scl_2002} or by applying a yaw torque which decays to zero \cite{naveh_dc_1999} as the unicycle's roll angle converges to zero.  The present paper considers the effect of periodic yaw oscillations on the open loop stability of a hydrodynamic Chaplygin sleigh with roll and shows that the system behaves as a parametric oscillator, which is a distinct result compared to the existing literature on the control of unicycle dynamics.

The rest of the paper is organized as follows. In \ref{sec:kinematic_model}  the kinematic model of a Chaplygin sleigh with roll dynamics is derived, and in \ref{sec:eom}, its equations of motion are derived. In \ref{sec:approx_roll}, the equation of approximate roll dynamics of the Chaplygin sleigh is derived by assuming a prescribed limit cycle in the reduced velocity space. In section \ref{sec:linearization}, the roll equation is linearized about the vertically upward equilibrium position, and in \ref{sec:mathieu}, the linearized equation is shown to be a nonhomogeneous damped Mathieu equation. The response of the homogeneous and nonhomogeneous Mathieu equation is analyzed in \ref{sec:lin_homogeneous} and \ref{sec:lin_nonhom}, respectively using Floquet theory, and stability charts are constructed in parameter space. In section \ref{sec:numerics}, a comparison is made of the solutions obtained via direct numerical simulation and analytical solutions from \ref{sec:lin_homogeneous} and \ref{sec:lin_nonhom}.

\section{Mathematical Model}
\subsection{Kinematics} \label{sec:kinematic_model}
\begin{figure}[htbp]\centering\includegraphics[width=1\columnwidth]{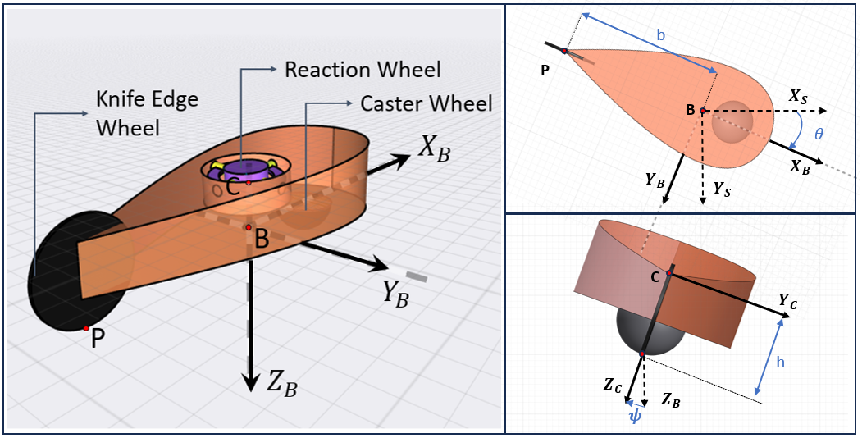}\caption{A Chaplygin sleigh with roll motion containing an internal rotor is shown on the left. In our idealized model (shown in the right two panels),  we assume a periodic torque along the body Z-axis.}
\label{fig:sleigh} \end{figure}

A conceptual model of a Chaplygin sleigh is shown in  fig.\ref{fig:sleigh}; it is an extruded body with a small inertia-less knife-edge wheel that is in contact with a rigid flat surface (the bottom of a pool) at point $P$. The sleigh is supported by a single caster  at the front that allows motion in any direction.   We assume that the sleigh is subject to a periodic torque in the body yaw direction. This is an approximation of the torque generated by the sinusoidal spin oscillations of an internal rotor whose angular acceleration can transfer an input torque on the sleigh. The sleigh is assumed to be on the bottom of a body of water pool while being completely submerged, and the water is modeled as an inviscid fluid. 

The configuration manifold of the physical system is $Q = SE(2) \times S^{1}$ parameterized locally by $q = (x, y, \theta, \psi)$, with generalized velocities $\dot{q} = T_{q}\in Q$ the tangent space to $q \in Q$. The spatial frame is denoted by $\mathcal{F}_S$ with axes $X_S-Y_S-Z_S$ and associated unit vectors $(\hat{i}, \hat{j}, \hat{k})$. The center of mass of the sleigh is at point $C$ with coordinates $(x_{b},y_{b} + h \sin{\psi},-h\cos{\psi} )$ in the spatial frame $\mathcal{F}_S$ and point $B$ is its projection on the ground with coordinates $(x_{\textcolor{red}{b}},y_{\textcolor{red}{b}},0)$. Additionally, the body frame collocated at $B$ and rotated by the yaw angle $\theta$ with respect to the spatial frame is denoted by $\mathcal{F}_B$ with axes $X_B-Y_B-Z_B$ and associated unit vectors $(\hat{i}_b, \hat{j}_b, \hat{k}_b)$; body frame attached at point $C$ and rotated by the roll angle $\psi$ with respect to the frame $B$ is denoted by $\mathcal{F}_C$ with axes $X_C-Y_C-Z_C$ and associated unit vectors $(\hat{i}_c, \hat{j}_c, \hat{k}_c)$. The coordinates of $P$ in frame $\mathcal{F}_B$ are $(-b,0,0)$. The $X_B$ axis is chosen to cross point $B$ and $P$ while the $Z_B$ axis is directed towards the ground so that clockwise rotation is positive. The rotation transformation $R_{BS}$ maps  vectors from frame  $\mathcal{F}_S$ to $\mathcal{F}_B$ and $R_{CB}$ maps vectors from frame $\mathcal{F}_B$ to $\mathcal{F}_C$,
\begin{equation*}\label{Rotmat}
    R_{BS} = \begin{pmatrix} 
        \cos\theta & \sin\theta & 0 \\ 
        -\sin\theta & \cos\theta & 0\\
        0 & 0 & 1
        \end{pmatrix} ~~,~~ 
        R_{CB} = \begin{pmatrix}
        1 & 0 & 0\\
        0 & \cos\psi & \sin\psi \\
        0 & -\sin\psi & \cos\psi
        \end{pmatrix}.
\end{equation*}
The angular velocity of the frame $\mathcal{F}_C$ with respect to the spatial frame is 
\begin{equation}
        \omega_{C} = \begin{bmatrix} \dot{\psi} \\ \dot{\theta}\sin\psi\\ \dot{\theta} \cos\psi \end{bmatrix} .\label{eq:wb}
\end{equation}

The velocity $(\dot{x}_{b}, \dot{y}_{b})$ in the spatial frame transform to $(u,v)$ in the $\mathcal{F}_B$ frame as,
\begin{equation}\label{eq:uv_xy}
V^B_B = \begin{bmatrix}
u \\ v \end{bmatrix} = \begin{bmatrix} 
        \cos\theta & \sin\theta \\ 
        -\sin\theta & \cos\theta
        \end{bmatrix} \cdot \begin{bmatrix}
            \dot{x}_{b}\\ \dot{y}_{b}
        \end{bmatrix}.
\end{equation}
\\
The velocity of the center of mass $C$ in body frame $\mathcal{F}_B$ is  $$V^B_C = V^B_B + \dot{s}^B_{CB} + \omega_B \times s^B_{BC},$$ where $\omega_B = \dot{\theta \hat{k}}$ is the angular velocity at point $B$ and $s^B_{CB} = h \sin(\psi) \hat{j} -h \cos(\psi) \hat{k}$ is the position of point $C$ from point $B$ in frame $\mathcal{F}_B$. Therefore velocity of the center of mass is
\begin{equation}
        V^{B}_{C} = \begin{bmatrix} u-h \dot{\theta}\sin\psi\\ v+h \dot{\psi}\cos\psi\\ h \dot{\psi}\sin\psi \end{bmatrix}.
\end{equation}\label{eq:vb}
We assume that the rear wheel at $P$ prevents slipping in the transverse $Y_B$ direction but rolls without any slipping in the longitudinal direction along the axis $X_B$.  While $dim(T_qQ) = 4$, the nonholonomic constraint at point $P$ given by 
 \begin{subequations}
  \begin{align}\label{NHC}
    V^{B}_{P}  = [-\sin\theta  ~\cos\theta ~-b ~~0] [\dot{x}_{b}~~\dot{y}_{b} ~~\dot{\theta}~\dot{\psi}~~0]^{\intercal} = 0 \nonumber \\
 \text{or} \quad v - b \dot{\theta}  = 0\\
   \intertext{with pfaffian one form being}
   (-\sin\theta) dx_{b} + (\cos\theta) dy_{b} - b d\theta = 0 \label{one_form}.
  \end{align}
\end{subequations}
The constraint restricts the generalized velocity $\dot{q} \in W(q) \subset T_qQ$,  where $dim(W(q)) = 3$. Therefore a reduced set of velocities $\bm{\eta} = [u,\dot{\theta}, \dot{\psi}]$ span the space $W(q)$ of allowable velocities. 

\subsection{Equations of Motion} \label{sec:eom}
The Lagrange function for the system is given by the difference in kinetic and potential energy $ \mathcal{L} = \mathcal{T-V}$, here potential energy ($\mathcal{V}$) of the system is given by the difference in work done by gravitational and buoyancy force,
\[\mathcal{V} = (m-\rho_{w}V_b)\cdot gh \cos\psi \]
$\rho_w$ is the density of water and $V_b$ is the volume of the body. The buoyancy force is the weight of the water displaced by the sleigh. The center of buoyancy and the center of gravity coincides with the center of mass as the sleigh is fully immersed in water and is also assumed to be of uniform density. The total kinetic energy ($\mathcal{T}$) is the sum of linear and rotational kinetic energies,
\[\mathcal{T} = \frac{1}{2}\Big((V^{B}_{C})^\intercal \cdot M\cdot V^{B}_{C} + \omega_{C}^\intercal \cdot I \cdot \omega_{C}\Big)\]
An additional hydrodynamic effect we consider in our model is that of added mass. In a physical sense, added mass is the inertia added to the system to compensate for the work done in changing the kinetic energy of the surrounding fluid as the body moves through it \cite{Lamb, milne-thomson96}. The sleigh has a mass $m$ and moment of inertia about the center of mass $I_c$. We assume our body to be prolate spheroid and hence has three planes of symmetry which makes the added mass and moment of inertia tensors diagonal. $M_a = diag(m_{11}, m_{22}, m_{22}),$ $I_a =  diag(m_{44}, m_{55}, m_{55})$. The transformed mass and inertia matrix are,
\begin{align*}
M =~~ mI_{3\times 3} + M_a &= 
    \begin{bmatrix} 
        m + m_{11} & 0 & 0\\ 0 & m + m_{22} & 0\\ 0 & 0 & m+m_{22}
    \end{bmatrix},\\ 
I =~~ I_c + I_a &= 
    \begin{bmatrix} 
        I_{{x}}+m_{44} & 0 & 0\\ 0 & I_{{y}} + m_{55} & 0\\ 0 & 0 & I_{{z}}+m_{55}
    \end{bmatrix}  
\end{align*}
When substituting velocities, mass, and moments of inertia in the Lagrange function, we get
\begin{equation}\label{eq:Lagrange}
    \begin{aligned}
        \mathcal{L} = &\frac{1}{2}\bigg[ \dot{\psi}^2 \Big(I_{x} + m_{44}+ (m+m_{22})h^2)\Big) + \Big(m+m_{11}\Big) \Big(u^2 -h\dot{\theta}u\sin\psi\Big) + \\ &\dot{\theta}^2 \Big(I_{y}\sin^2\psi + I_{z}\cos^2\psi + (m+m_{11})h^2 \sin^2\psi + m_{55}\Big) +\\ &\Big(m+m_{22}\Big) \Big(v^2 -h\dot{\psi}v\cos\psi\Big) - (m - \rho_w V_b)gh\cdot\cos\psi\bigg]
    \end{aligned}
\end{equation}

The viscous dissipation in the system is 
\[
R = \frac{1}{2}c_1(\dot{x}^2 + \dot{y}^2) + \frac{1}{2}c_2 \dot{\theta}^2 + \frac{1}{2}c_{\psi} \dot{\psi}^2.
\]
The Euler-Lagrange equations are given by
\begin{equation}\label{eq:Euler_Lag}
  \frac{\mathrm{d}}{\mathrm{d}t}\Big(\frac{\partial \mathcal{L}}{\partial \dot{q}^k }\Big)-\frac{\partial \mathcal{L}}{\partial q^k}=\lambda C_k -\frac{\partial{R}}{\partial{q^k}} + \Gamma^k 
\end{equation}
where, $\Gamma^k$ are the generalized forces, $\lambda$ is Lagrange multiplier and $C_{k}$ is the coefficient of the one form $dq_{k}$ (\ref{one_form}).

Straightforward calculations on \eqref{eq:Euler_Lag} yield four second-order differential equations, which can be transformed into eight first-order equations. Here we make the observation that the Lagrangian is independent of $x$, $y$, and $\theta$, and therefore the accelerations $\ddot{x}$, $\ddot{y}$ and $\ddot{\theta}$ are independent of $x$, $y$ and $\theta$. Using the nonholonomic constraint \eqref{NHC}, eliminating the Lagrange multiplier and transforming to the body velocities $(u,v)$ using \eqref{eq:uv_xy}, we can reduce the number of equations of motions by one. The dissipation function can also be rewritten in body frame velocities using the transformations $\dot{x} =u\cos{\theta} - v\sin{\theta}$ and $\dot{y} = u\sin{\theta} + v\cos{\theta}$ and the nonholonomic constraint $v = b\dot{\theta}$ as
\begin{equation}\label{eq:R}
R = \frac{1}{2}C_uu^2 + \frac{1}{2}C_{\theta} \dot{\theta}^2 + \frac{1}{2}C_{\psi} \dot{\psi}^2.
\end{equation}

The reduced equations can be written in the form
\begin{equation}\label{eq:full}
    \mathcal{M}(\psi)\bm{\dot{\eta}} + \mathcal{C}(\psi,\dot{\psi},\dot{\theta})\bm{\eta} + \mathcal{G}(\psi) = \Gamma 
\end{equation}
where $\bm{\eta} = [u,\dot{\theta}, \dot{\psi}]$ is the reduced generalized velocity, $\mathcal{M}$ the mass matrix,  $\mathcal{C}$ the Coriolis and Centrifugal terms and  $\mathcal{G}$ the gravitational term. We assume that the sleigh is subject to a periodic torque $ A\sin{\Omega t}$ (frequency $\Omega$  and time period $T = \frac{2\pi}{\Omega}$) aligned with the body $z-$ axis in frame $\mathcal{F}_C$.  The terms in the matrices $\mathcal{M}$ $\mathcal{C}$, $\mathcal{G}$ and $\Gamma$ are
\begin{align*}
     \mathcal{M}_{11} & = -m-m_{11} ~ ,~\mathcal{M}_{12} = \mathcal{M}_{21} = (m+m_{11})\cdot h\sin\psi \\ \mathcal{M}_{22} &= -\Big((m+m_{11})h^2\sin^2\psi +(m+m_{22})b^2 + I_{y}\sin^2\psi  + I_{z}\cos^2\psi + m_{55}\Big)\\ \mathcal{M}_{23} &= \mathcal{M}_{32} = -(m+m_{22})bh \cos\psi~,~ \mathcal{M}_{33} = -\Big(I_{x} + m_{44} + (m+m_{22})h^2\Big)\\
     \mathcal{C}_{11} &= -C_{u} ~~,~~\mathcal{C}_{12} = (m+m_{22})b\dot{\theta} + (2m+m_{11}+m_{22})h\dot{\psi}\cos\psi\\ \mathcal{C}_{21} &= (m_{11}-m_{22})h\dot{\psi}\cos\psi - (m+m_{22})b\dot{\theta} ~,~\mathcal{C}_{23} = (m+m_{22})bh \dot{\psi}\sin\psi \\\mathcal{C}_{22} &= -C_{\theta}-\dot{\psi}\sin2\psi((m+m_{11})h^2+I_{y}-I_{z}) ~,~\mathcal{C}_{31} = -(m+m_{11})h\dot{\theta}\cos\psi \\ \mathcal{C}_{32} &= \frac{\dot{\theta}\sin2\psi((m+m_{11})h^2+I_{y}-I_{z})}{2}~,~ \mathcal{C}_{33} = -C_{\psi} \\ \Gamma &= [0~ -A \sin \Omega t \cos{\psi} ~~ 0]^{\intercal}~~,~~ \mathcal{G} = [0~~0~~(m-\rho_w V_b)gh\sin\psi]^{\intercal}.
    \end{align*}

\subsection{Approximation of parametric roll motion} \label{sec:approx_roll}
When a planar Chaplygin sleigh (therefore one without any roll motion) with viscous dissipation is subjected to a sinusoidal torque with period $T$, the velocity $u$ and the yaw angular velocity $\dot{\theta}$ converge to a limit cycle in the reduced velocity space spanned by $(u,\dot{\theta})$ see, for instance,  \cite{ft_nody_2018, borisov_nody_2019}.  Furthermore, numerical simulations showed that $u$ and $\dot{\theta}$ are $kT$-periodic, and their solution converges to a limit cycle with a figure-8 shape, which implies that the frequency of oscillation in $u$ is double the frequency of oscillations in $\dot{\theta}$. Using the harmonic balance method, an approximate solution for $u$ and $\dot{\theta}$ were constructed in \cite{ft_nody_2018}, showing that $u$ has only second harmonics, while yaw rate $\dot{\theta}$ has only first harmonic with zero means. 
 
\begin{figure}[htbp]
\centering\includegraphics[width=1\columnwidth]{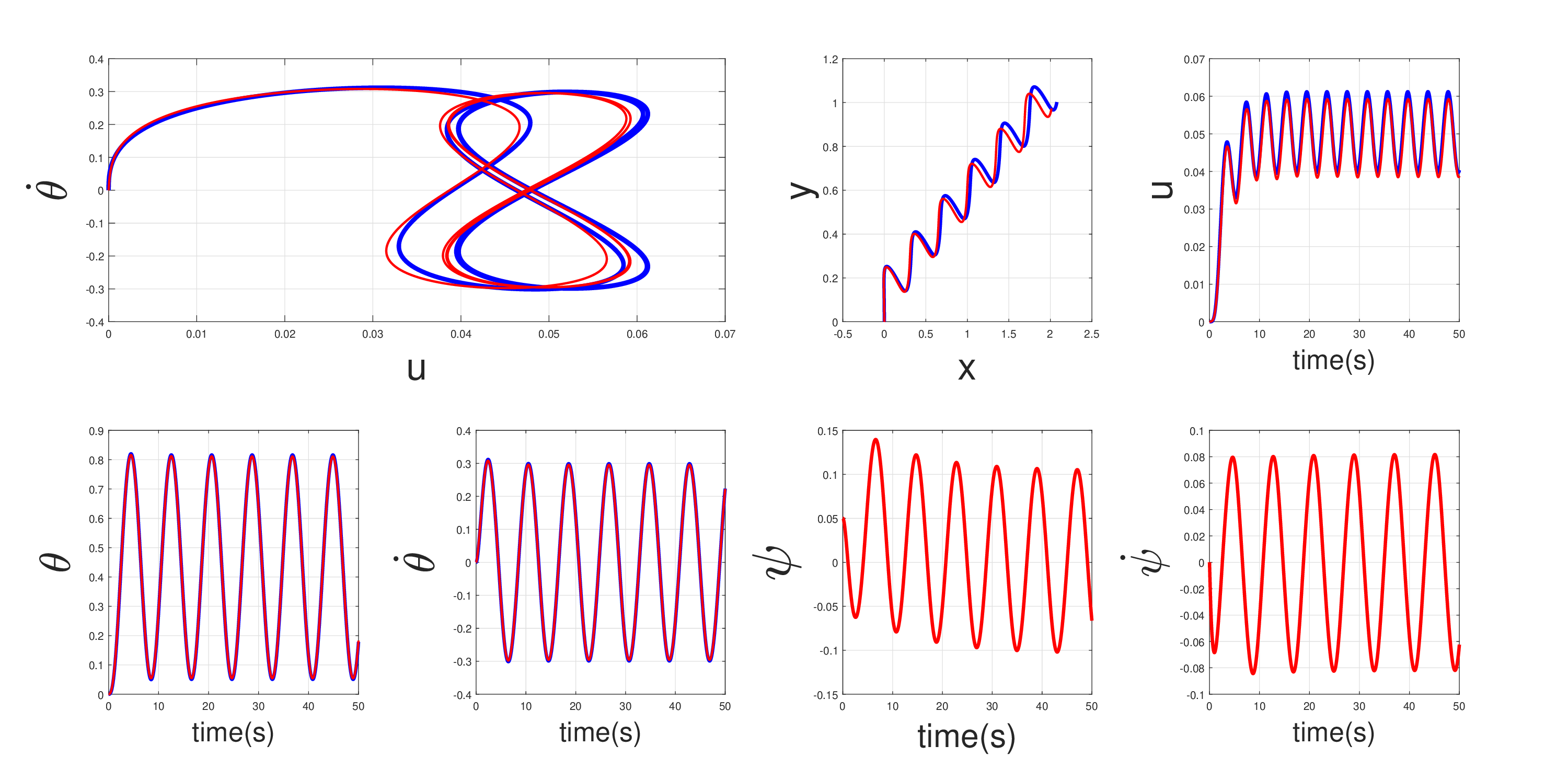}\caption{\textcolor{blue}{Blue Trajectory}: Numerical simulation of a planar hydrodynamic Chaplygin sleigh, \textcolor{red}{Red Trajectory}: Numerical simulation of hydrodynamic Chaplygin sleigh with roll dynamics. Sinusoidal force input in both the systems has an amplitude of $A = 0.103$ and frequency as $\Omega = 0.77$}\label{fig:sample_sim} 
\end{figure}

Numerical simulations show that such limit cycles persist even when the sleigh possesses hydrodynamic added mass. A limit cycle shaped as a figure-8 is shown (in blue) fig.\ref{fig:sample_sim}(a)  for the case when a planar Chaplygin sleigh with dissipation and hydrodynamic added mass is given a periodic torque, while the other subfigures in fig.\ref{fig:sample_sim} showing the evolution of the configuration variables and velocities. The same figures show (in red) the respective velocities and configuration variables when the Chaplygin sleigh also has roll motion. It can be seen that the $u$ and $\dot{\theta}$ are largely unaffected by the roll motion, but the roll angle $\psi$ and roll velocity $\dot{\psi}$ are affected by the periodic velocity $u$ and yaw rate $\dot{\theta}$. 

We therefore use the approximation that $u$ and $\dot{\theta}$ are uncoupled from the roll dynamics of the sleigh, but they affect the roll motion. This partial decoupling means that $u$ and $\dot{\theta}$ can be approximated as similar to the solution in \cite{ft_nody_2018} but including the effect of the hydrodynamic added mass.
\begin{equation}\label{eq:lc}
    \begin{aligned}
        u_p & = u_c + \alpha_s \sin(2\Omega t)+ \alpha_c \cos(2\Omega t)\\ \dot{\theta_p} &= \beta_s \sin(\Omega t) + \beta_c \cos(\Omega t). 
    \end{aligned}
\end{equation}
In \eqref{eq:lc} the parameters $u_c, ~\alpha_s,~\alpha_c, ~\beta_s$ and $\beta_c$ can be determined by direct substitution of the assumed solution \eqref{eq:lc} in \eqref{eq:full} while approximating $(\psi \approx 0, h = 0)$. This leads to the following non-linear equations  where we denote $\sigma_4 = I_{z} +m_{55} +b^2 (m+m_{22})$.
\begin{subequations}
   \begin{flalign}
         0 = &~b{(m+m_{22})}{({\beta_c }^2 +{\beta_s }^2)}-2C_u u_c \label{parameter_eqn_a}\\
        - 4 \Omega \alpha_c {(m+m_{11})} =  &~2b \beta_c \beta_s {(m+m_{22}) -2C_u \alpha_s} \label{parameter_eqn_b}\\
        4 \Omega \alpha_s {(m+m_{11})} = &~b{(m+m_{22} )}{({\beta_c }^2 -{\beta_s }^2 )}-2C_u \alpha_c \label{parameter_eqn_c}\\
         - 2\Omega \beta_c \sigma_4 = &~2A-2C_{\theta} \beta_s -b{(m+m_{22} )}{(\alpha_s \beta_c -\alpha_c \beta_s +2\beta_s u_c )} \label{parameter_eqn_d}\\
        \Omega \beta_s \sigma_4 =  &~-2C_{\theta} \beta_c -b{(m+m_{22} )}{(\alpha_c \beta_c +\alpha_s \beta_s +2\beta_c u_c )}\label{parameter_eqn_e}
   \end{flalign}\label{eq:parameter_eqn}
\end{subequations}

The nominal limit cycle solution of $u$ and $\dot{\theta}$ can then be used to obtain a single second-order differential equation describing the roll motion of the Chaplygin sleigh, which is of the form, 
\begin{equation}\label{eq:psidot}
    \ddot{\psi} = F(u(t), \dot{\theta}(t), \psi, \dot{\psi}) = {F}(\psi,\dot{\psi}, t; A,\Omega,u_c,\alpha_s, \alpha_c, \beta_s, \beta_c)
\end{equation}
where the substitution of the limit cycle solution for $u(t)$ and $\dot{\theta(t)}$ leaves the function $F$ dependent on the parameters $[A,\Omega,u_c,\alpha_s, \alpha_c, \beta_s, \beta_c]$ associated with the limit cycle \eqref{eq:lc}.

\section{Roll motion of the Chaplygin sleigh - Linear Analysis} \label{sec:linear}
\subsection{Linearization}\label{sec:linearization}
Simulations such as those shown in fig.\ref{fig:sample_sim} suggest that the hydrodynamic Chaplygin sleigh with roll motion has a topologically similar limit cycle in the reduced velocity space as that of planar Chaplygin sleigh while the roll angle is in the neighborhood of its upper equilibrium where $(\psi_{e}=0)$. Therefore, to analyze the parameters that affect the stability of roll angle at this equilibrium, we linearize the nonlinear vector field $F$ in \eqref{eq:psidot} about the equilibrium position. 

\begin{equation}\label{eq:linear}
    \ddot{\psi} \approx {F_{\psi}}(\psi_{e} ,\dot{\psi_e})+\begin{bmatrix} \frac{\partial {F_{\psi}}}{\partial \psi } & \frac{\partial {F_{\psi}}}{\partial \dot{\psi}} \end{bmatrix}\biggr\rvert_{(\psi_{e} ,\dot{\psi_e})} \cdot \begin{pmatrix} \psi -\psi_e \\ \dot{\psi} -\dot{\psi_e} \end{pmatrix} + \mathcal{O}(\psi_{e} ,\dot{\psi_e})^{2}
\end{equation}
The equilibrium point we want to analyze for stability is the upright position ($[\psi_{e},\dot{\psi_e}] = [0,0]$) of the modified hydrodynamic Chaplygin sleigh. By algebraic manipulation and simplification of (\ref{eq:linear}) we get a non-homogeneous linear second-order differential equation with periodic coefficients given as
\begin{equation}\label{eq:lin_dim}
    \ddot{\psi} + \Big(\delta_m + \epsilon_m\cdot \cos (2\Omega t - \gamma_m) \Big)\psi + \xi_m(t) \cdot\dot{\psi} = C(t)
\end{equation}
Where $\sqrt{\delta_m}$ is the natural frequency of the system, and $\epsilon_m$ is the amplitude of the parametric excitation. If we re-scale time as $\tau = 2\Omega t$, and $\frac{\partial^2 \psi }{\partial t^2} = 4\Omega^2\frac{\partial^2\psi}{\partial \tau^2}$. The linearized governing equation for the roll motion of the modified hydrodynamic Chaplygin sleigh (\ref{eq:lin_dim}) is non-dimensionalized and can be written as
\begin{equation}\label{eq:non_hom_lin} 
    \ddot{\psi} + \Big(\delta + \epsilon\cdot \cos(\tau - \gamma)\Big)\psi + \xi(\tau) \cdot\dot{\psi} = C(\tau) 
\end{equation} 
The natural frequency of the linearized parametric oscillator is 
\begin{align}\label{eq:delta} 
    \begin{split} 
        \delta ~=& \frac{-1}{4 \Omega^2 \sigma_1}\Bigg(\Big((m+m_{22})b^2 + I_{z} + m_{55}\Big)(m-\rho_w V_b)gh  \\ &~~~+(m+m_{22})C_{u} u_{c}bh^2 + \sigma_3\Big(\frac{\beta_{c}^2 + \beta_{s}^2}{2}\Big)\Bigg) 
    \end{split}
\end{align}
The small parameter $\epsilon$ is
\begin{align}\label{eq:epsilon}
    \epsilon ~=  \frac{\sqrt{\gamma_{c}^2 + \gamma_{s}^2}}{4 \Omega^2 \sigma_1}~~,~~
    \gamma~ = & \arccos{\Big(\frac{\gamma_c}{\sqrt{\gamma_{c}^2 + \gamma_{s}^2}}\Big)}  
\end{align}
where
\begin{flalign*} 
    \sigma_1 =& (I_{z}+m_{55})\big((m+m_{22})h^2 + I_{x} + m_{44}\big) + (I_{x}+m_{44})(m+m_{22})b^2 \\ 
    \sigma_2 =& \Big((I_{z}+m_{55})(m+m_{11})+(m+m_{22})(m_{11}-m_{22})b^2\Big)\\ 
    \sigma_3 =& (I_{y}-I_{z})\big((m+m_{22})b^2 + I_{z} + m_{55}\big) + \sigma_2 h^2\\
    \gamma_c =& -\Big(\sigma_3 (\frac{\beta_{c}^2 - \beta_{s}^2}{2}) + C_{u}(m+m_{22})bh^2\alpha_{c}\Big)\\
    \gamma_s = &-\big(\sigma_3 \beta_{s}\beta_{c} +C_{u}(m+m_{22})bh^2\alpha_{s}\big).
\end{flalign*}
The effective damping coefficient is
\begin{align}\label{eq:damping}
    \begin{split}
        \xi(\tau) ~=& \frac{1}{2 \Omega \sigma_1}\Bigg((m+m_{22})(m_{11}-m_{22})bh^2 (u_c + \alpha_c \cos\tau + \alpha_s\sin\tau) \\ & + C_{\psi}\big((m+m_{22})b^2 + I_z+m_{55}\big)\Bigg).
    \end{split}
\end{align}

The natural frequency $\delta$, damping coefficient $\xi$, and parametric amplitude $\epsilon$ all depends on the parameters of the limit cycle; in particular, the amplitude $\sqrt{\beta_c^2+\beta_s^2}$ of the limit cycle  $\dot{\theta}(t)$, $\alpha_c$ and $\alpha_s$ that determine the amplitude of the limit cycle solution $u(t)$ and also the added mass parameters $m_{11}$, $m_{22}$ and $m_{55}$. The damping coefficient $\xi$ also varies periodically with a frequency equal to the fundamental frequency of the limit cycle in \eqref{eq:lc}. The natural frequency also depends on the drift velocity $u_c$ and decreases with the \color{blue}increase \color{black} in drift velocity. The natural frequency is positive only if $(m- \rho_w V_b) < 0$, i.e., when the body is lighter than water. This is, however not a sufficient condition and the natural frequency may still be negative depending on the relative magnitudes of the amplitude of limit cycles, the mean value of $u$ on the limit cycle, and the added mass. The non-homogeneous (or forcing) term in \eqref{eq:non_hom_lin} arises as a result of linearizing the roll equation about a limit cycle solution of the planar Chaplygin sleigh with equilibrium approximation $(\psi_{e}\approx 0, \dot{\psi}_e \approx 0)$.
\begin{align}\label{eq:nonhom}
    \begin{split} 
        C(\tau&)=\frac{-h}{8\Omega^2\sigma_1}\Bigg(\Big(\sigma_2(2\beta_c u_c +\alpha_s\beta_s+\alpha_c\beta_c) -2C_{\theta}(m+m_{22}) b \beta_{c}\Big)\cos(\frac{\tau}{2}) \\ & + \sigma_2 \Big(\alpha_c\beta_c-\alpha_s\beta_s\Big) \cos(\frac{3}{2}\tau)+ \sigma_2 \Big(\alpha_s\beta_c +\alpha_c\beta_s\Big)\sin(\frac{3}{2}\tau)\\ &+\Big(2b(A -C_{\theta} \beta_{s})(m+m_{22})+~\sigma_2 (2\beta_s u_c+\alpha_s\beta_c -\alpha_c\beta_s)\Big)\sin(\frac{\tau }{2})\Bigg).
    \end{split}
\end{align}

\subsection{The Mathieu oscillator approximation} \label{sec:mathieu}
When the elements of the added mass tensor $m_{11} = m_{22}$, and the hydrodynamic effects other than buoyancy are negligible or when the shape of the Chaplygin sleigh is almost a sphere. We can ignore the added mass tensor in our initial analysis. In this case the damping coefficient $\xi = \xi_c$ becomes constant,
\begin{equation}\label{eq:cosntant_damping}
    \xi_c = \frac{C_{\psi}(mb^2 + I_{C_z})}{2 \Omega \big(I_{C_z} (mh^2 + I_{C_x}) + I_{C_x}mb^2\big) }
\end{equation} 
and the linearized equation is then,
\begin{equation}\label{eq:mathieu} 
    \ddot{\psi} + \xi_c \cdot \dot{\psi} + \Big(\delta+\epsilon \cdot \cos(\tau - \gamma)\Big)\psi  = C(\tau)
\end{equation}
The homogeneous component of the linear equation is a Mathieu-type oscillator. The well-known Mathieu equation \eqref{eq:mathieu} has been extensively studied and is seen in the stability studies of periodic motions in nonlinear autonomous systems similar to the one in this paper. Floquet-Lyapunov theory, perturbation theory, and Hill's method of infinite determinants can be applied for stability analysis of linear homogeneous equations with periodic coefficients \cite{Mathieu_RRand,nayfeh2008perturbation,yakubovich_1975,Stroker_nonlinearvibr,Magnus_Winkler_Hillseqn}. We ignore the non-homogeneous term $C(\tau)$ to analyze stability of oscillatory solutions of only the homogeneous Mathieu equation using Floquet theory.

\textbf{Theorem 1} \cite{yakubovich_1975} \textit{(Floquet-Lyapunov theorem) - Any state-transition matrix $\mathbf{\Psi}(t)$ of equation \eqref{eq:mathieu} with $T$-periodic coefficients is expressible in the form of $$\mathbf{\Psi}(t) = P(t)e^{\mathbf{K}t}$$ where $P(t)$ is a non-singular continuous $T$-periodic $n\times n$ matrix function whose derivative is an integrable piece-wise continuous function and such that $P(0) = \mathbf{I}_n$, also  $\mathbf{K} =ln[\mathbf{\Psi}(T)]/T$ is some constant $n\times n$ matrix.}

For any initial condition, the solution for the equation \eqref{eq:mathieu} can be expressed using the state transition matrix, $\eta(t) = \mathbf{\Psi}(t,t_0)\eta(t_0)$. The state transition matrix evaluated for time period $T$ is known as the Monodromy matrix, $\mathbf{M} = \mathbf{\Psi}(T,0)$. Then for all $t\geq 0$ and  $k\in \mathbf{Z}^{+}$ we can obtain solution for equation \eqref{eq:mathieu} as
\[\eta(t) = \mathbf{\Psi}(kT + \tau,kT) \mathbf{M}^{k} \eta(0).\]

Hence the stability of the orbits of the homogeneous system \eqref{eq:mathieu} depends on the eigenvalues of Monodromy matrix $\mathbf{M}$ known as Floquet or characteristic multipliers $\lambda(\mathbf{M})$ see \cite{yakubovich_1975,Magnus_Winkler_Hillseqn}. The Mathieu equation \eqref{eq:mathieu} has an asymptotically stable solution if all of the multipliers lie inside the unit circle,  that is if $\lvert \lambda(\mathbf{M}) \rvert < 1$, and it is bounded if the multipliers are semi-simple eigenvalues with $\lvert \lambda(\mathbf{M}) \rvert = 1$. The solutions diverge if one or more of the eigenvalues of matrix $\mathbf{M}$ lie outside the unit circle ($\lvert \lambda(\mathbf{M}) \rvert > 1$) or if on the unit circle eigenvalues are not semi-simple.
 
To construct a stability chart in two-parameter ($\delta-\epsilon$) space for the Mathieu oscillator \eqref{eq:mathieu}; we first calculate the value of the parameters $[A,~\Omega,~u_c,~\alpha_s,~\alpha_c,~\beta_s,~\beta_c]$ for each point in the $\delta-\epsilon$ space by solving the set of seven non-linear equations \eqref{eq:parameter_eqn}(a-e), \eqref{eq:delta} and \eqref{eq:epsilon}. These equations have a unique solution for real and positive values of parameters $[A,~\Omega]$. Then using these planar limit cycle parameters we numerically simulate the system \eqref{eq:mathieu} with arbitrary initial conditions for period $T$ and obtain the Monodromy matrix and Floquet multipliers.
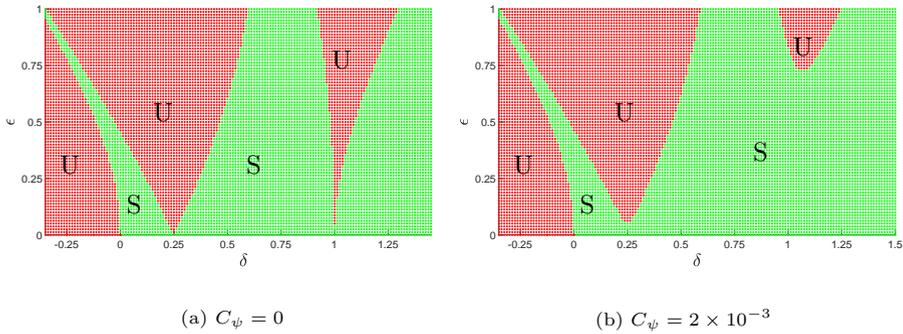
\begin{figure}
         \begin{picture}(360,100)\put(0,0){\includegraphics[width=0.49\columnwidth]{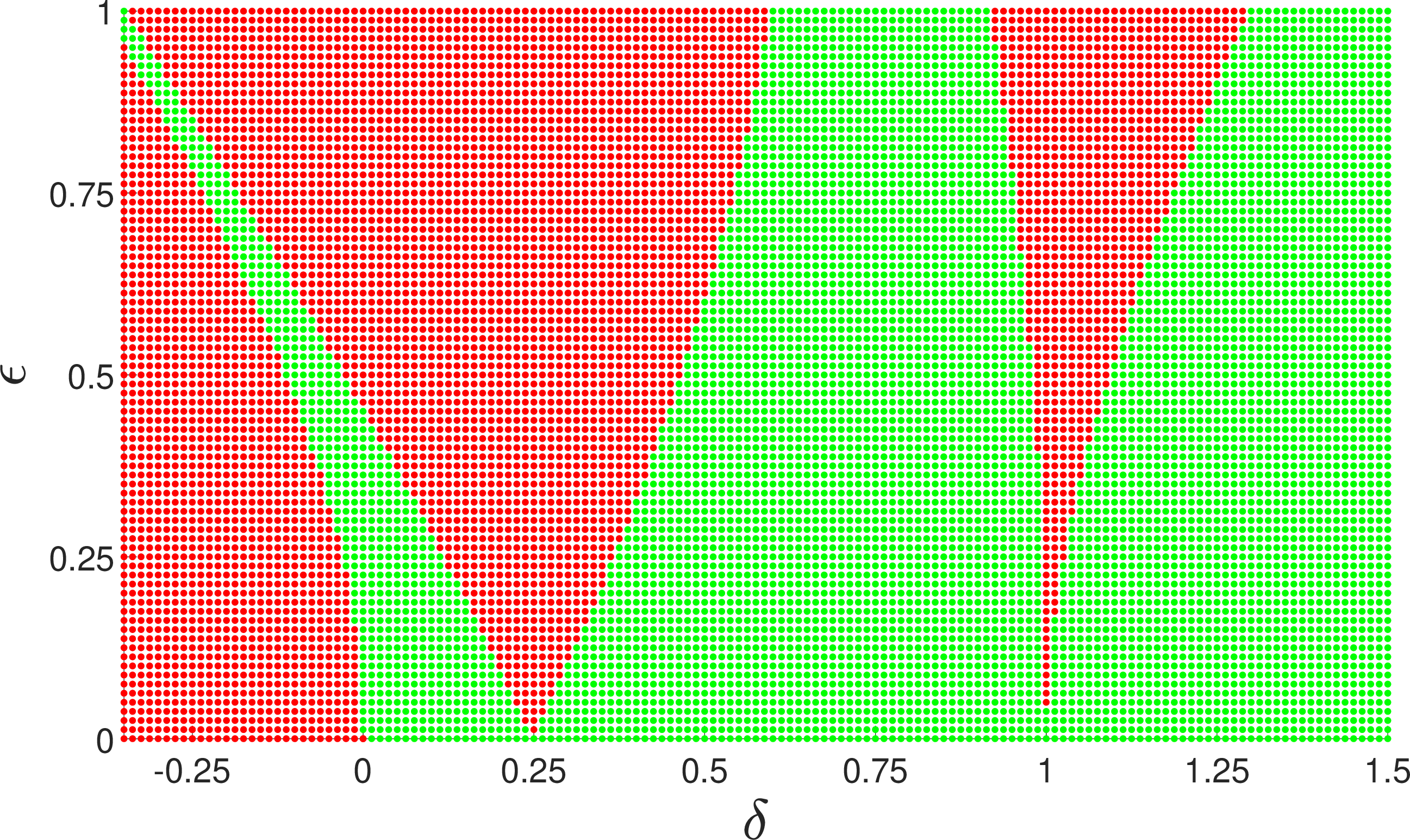}}\put(170,0){\includegraphics[width=0.49\columnwidth]{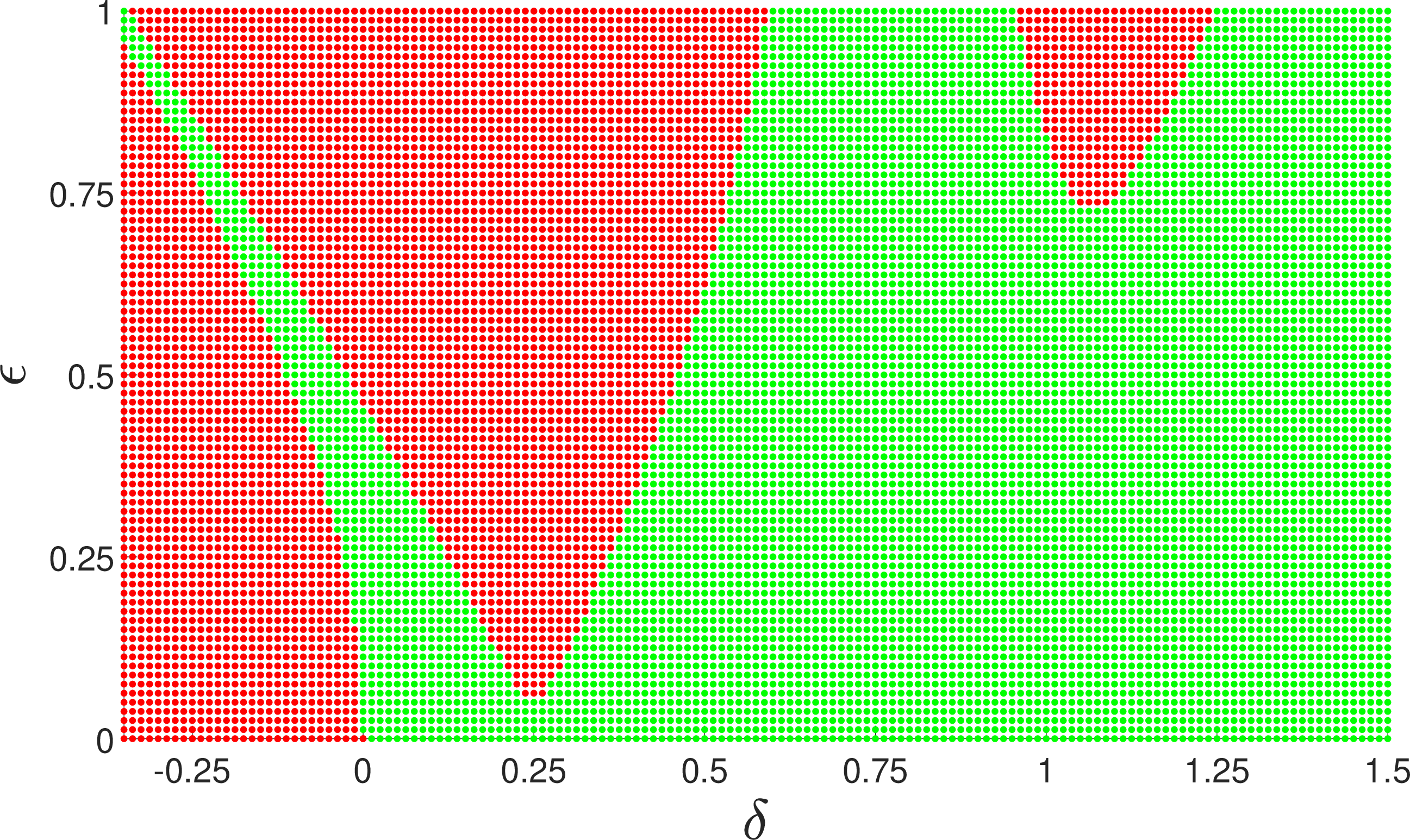}}\put(190,35){U}\put(20,35){U}\put(45,20){S} \put(215,20){S} \put(55,55){U} \put(228,55){U} \put(90,35){S} \put(280,40){S} \put(122,75){U} \put(295,80){U}\end{picture}
         \subfloat[$C_{\psi} = 0 $]{\hspace{.5\linewidth}}
         \subfloat[$C_{\psi} = 2\times10^{-3}$]{\hspace{.5\linewidth}}
        \caption{Stability chart for the Mathieu equation \eqref{eq:mathieu} in $\delta - \epsilon$ parameter space numerically obtained using Floquet theory.}
        \label{fig:mathieu_chart}
\end{figure}

The Floquet multipliers tell us about the stability of the system for the chosen parameters. Stability conditions on characteristic multipliers obtained from $Floquet-Lyapunov $ theorem are used to construct stable-unstable regions in the two-parameter space $(\delta-\epsilon)$  shown in fig.(\ref{fig:mathieu_chart}). The green region (labeled as $\textbf{S}$) is where the parameter values produce stable roll motion around the equilibrium point, and the red region (labeled as $\textbf{U}$) represents the unstable region. The instability is caused due to parametric resonance, which results from the frequency of parameter variation and the natural frequency of the system. The curves defined by the stable-unstable region in the Mathieu chart starting at $\delta = n^2/4, ~n=0,1,2,...$ and $\epsilon = 0$ have one of the Floquet multipliers equal to 1 which corresponds to $T$ periodic solutions. The curves starting at $\delta = (2n+1)^2/4$ and $\epsilon = 0$ have $\lambda(\mathbf{M}) = -1$ which corresponds to $2T$ periodic solutions.

\subsection{Linear homogeneous parametric oscillator with periodic damping} \label{sec:lin_homogeneous}
If we consider the effects of added mass in the linear system \eqref{eq:non_hom_lin} and consider first only the homogeneous equation ignoring the effects of the direct forcing term $C(\tau)$ we can still represent the linear parametric oscillator as $\dot{\eta} = A\eta$ where $A$ is $T$-periodic.
\begin{equation}\label{eq:hom_lin} 
\begin{bmatrix}\dot{\psi} \\ \ddot{\psi} \end{bmatrix} = \begin{bmatrix} 0 & ~~1 \\ -\Big(\delta + \epsilon \cdot \cos(\tau-\gamma)\Big) & ~~\xi(\tau) \end{bmatrix}  \begin{bmatrix}\psi \\ \dot{\psi} \end{bmatrix}
\end{equation}
The presence of added mass coefficient activates the time-periodic term in the damping coefficient $\xi(\tau)$ \eqref{eq:damping}. Linear parametric oscillators with time-periodic damping coefficient \eqref{eq:hom_lin} have been analyzed for their stability using the strained parameter method and Floquet theory in \cite{Burgh2002AnEW,Afzali2017AnalysisOT} where it is observed that a stability chart different from the Mathieu stability chart is possible due to the presence of periodic damping.
\begin{figure}[htbp]
\begin{picture}(360,60)
\put(-5,0){\includegraphics[width=0.3\columnwidth]{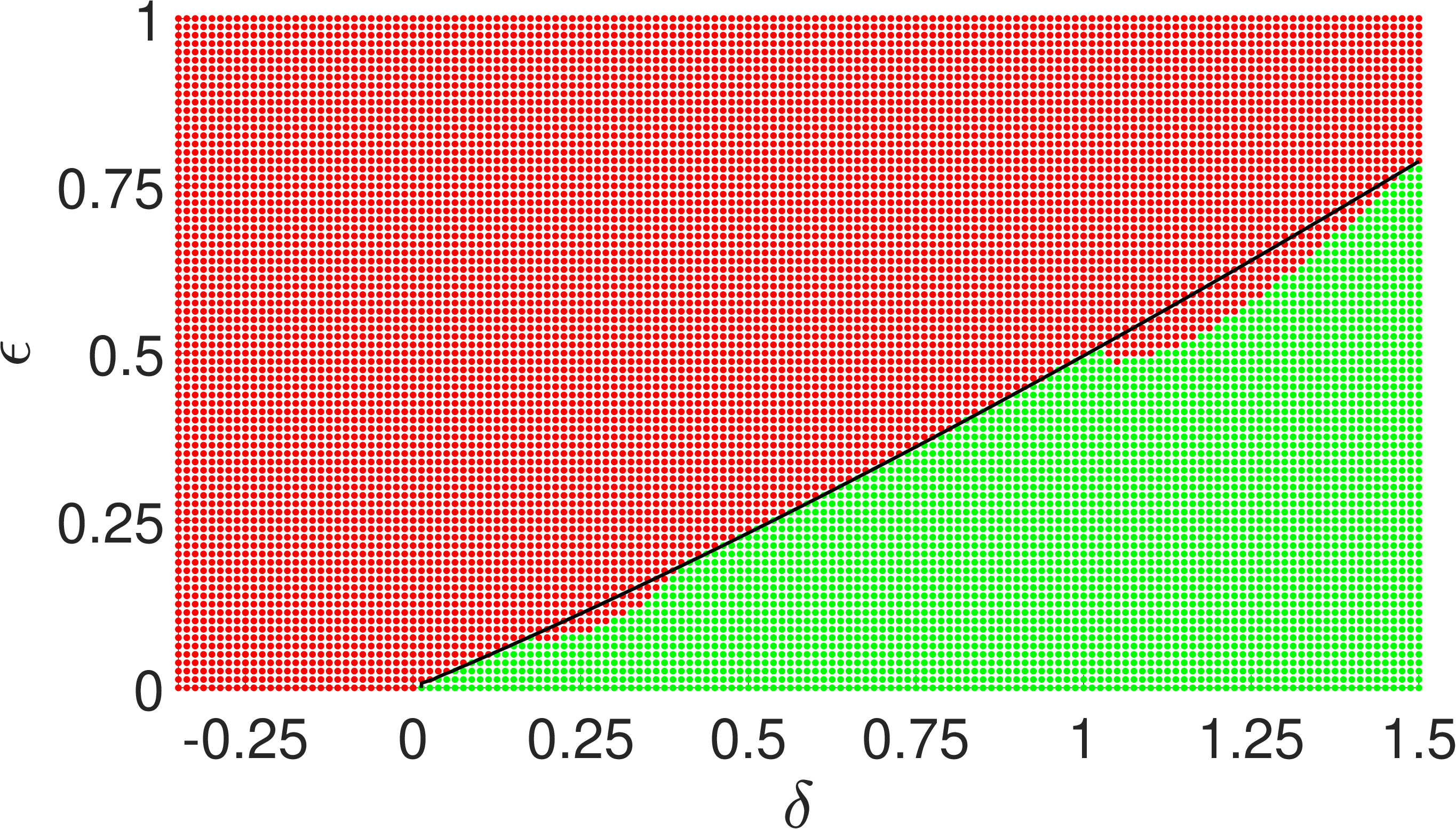}}
\put(115,0){\includegraphics[width=0.3\columnwidth]{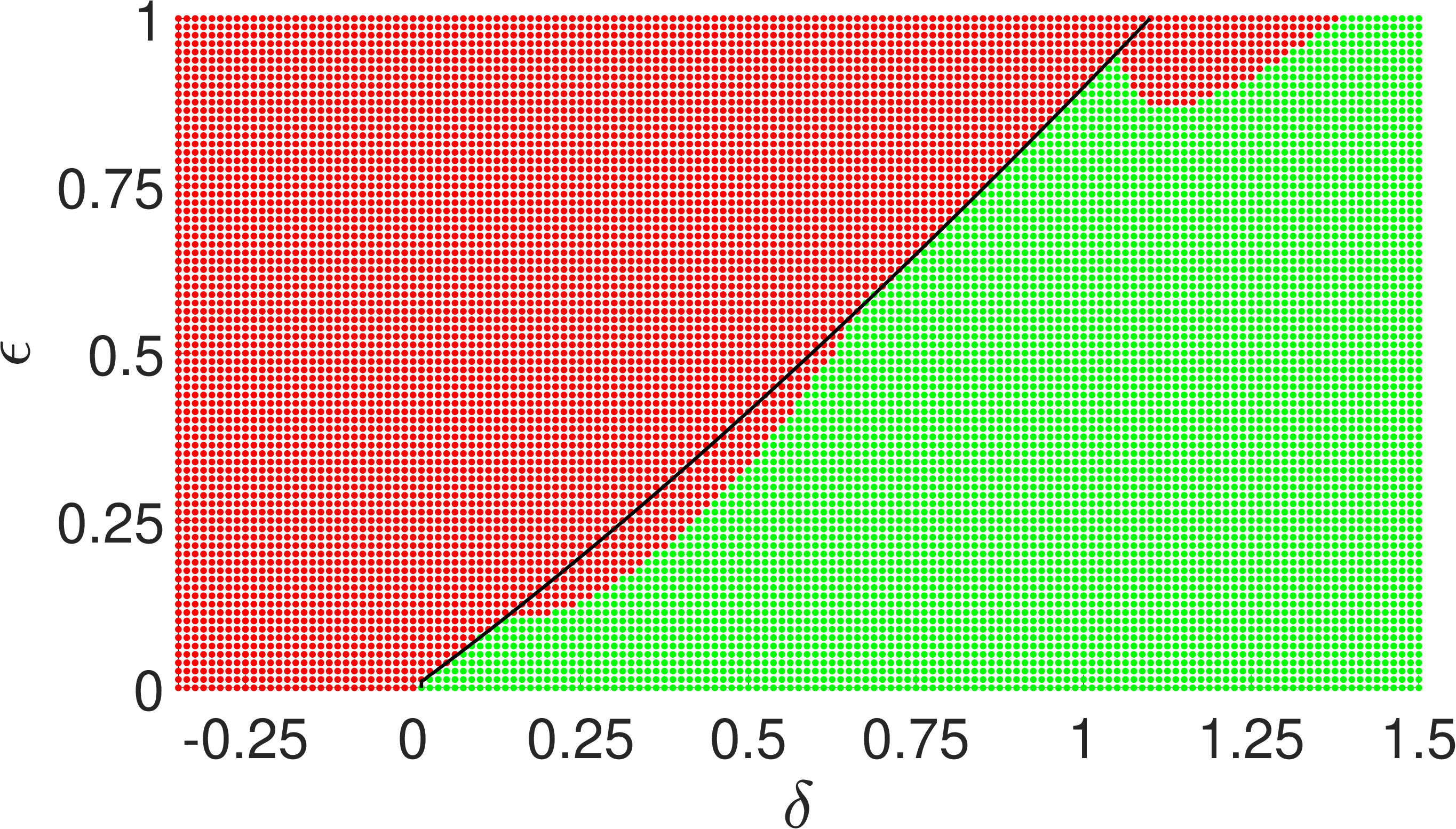}}
\put(235,0){\includegraphics[width=0.3\columnwidth]{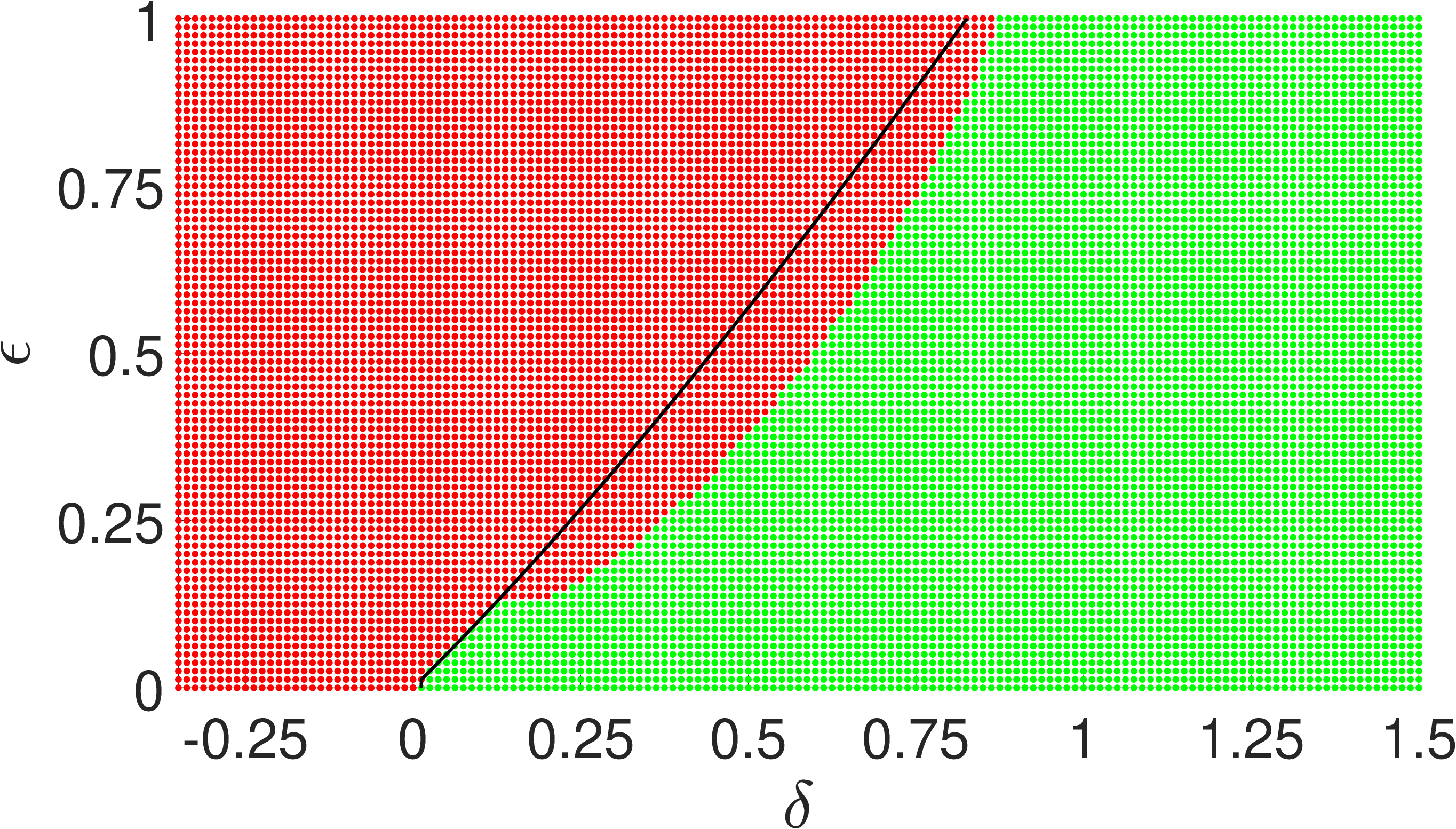}}\put(25,40){U} \put(145,40){U}\put(265,40){U} \put(75,20){S}\put(190,20){S}\put(310,20){S}
\end{picture}
\subfloat[$C_{\psi}= 7\times10^{-3}$]{\hspace{.3\linewidth}}
\subfloat[$C_{\psi} = 9\times10^{-3}$]{\hspace{.4\linewidth}}
\subfloat[$C_{\psi} = 1\times10^{-2}$]{\hspace{.35\linewidth}}
\caption{Stability chart for the linear equation \eqref{eq:hom_lin} in $\delta - \epsilon$ parameter space numerically obtained using Floquet theory.}\label{fig:stability_A}\end{figure}
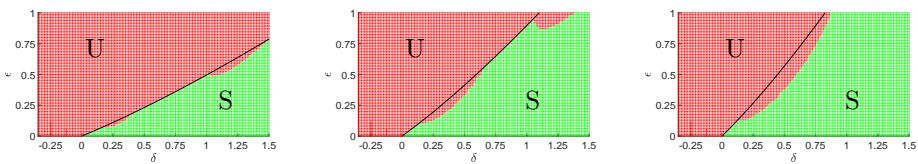

We again use Floquet theory to construct a stability chart for the system \eqref{eq:hom_lin} in the two-parameter space ($\delta - \epsilon$) as shown in fig.(\ref{fig:stability_A}). The parameter range where the characteristic multiplier is greater than 1 is shown in red (and labeled by \textbf{U}). The stability chart fig.(\ref{fig:stability_A}) is qualitatively different from the stability chart in fig.(\ref{fig:mathieu_chart}) due to the periodic damping. The damping coefficient $\xi(t)$ given in \eqref{eq:damping} can be rewritten compactly as  
\begin{equation} \label{eq:compact_damping}
    \xi(\tau) = \frac{1}{\Omega}\Big(\xi_1 C_{\psi}  + \xi_2 (u_c +\alpha_c \cos \tau + \alpha_s \sin \tau)\Big)
\end{equation} 
where constants $\xi_1 = \frac{((m+m_{22})b^2+I_{C_z}+m_{55})}{2\sigma_1}$,  $\xi_2 =\frac{(m+m_{22})(m_{11}-m_{22})bh^2}{2\sigma_1}$ and parameters $ \alpha_c , \alpha_s < u_c$.
 The time-averaged damping is 
\begin{equation}\label{eq:avg_damping}
    \overline{\xi} = \frac{1}{T}\int_0^T \xi d\tau = \frac{1}{\Omega } (\xi_1C_{\psi} + \xi_2 u_c).
\end{equation}
When the added mass coefficients $m_{11} < m_{22}$ (which is the case for prolate spheroid or other slender body shapes), the term $\xi_2<0$; this combined with sufficiently high values of average longitudinal velocity ($u_c$) will make the average damping $\overline{\xi}$ negative. The black curve in the fig.(\ref{fig:stability_A}), which divides the graph into two sections, represents the locus of $\overline{\xi} = 0$; to the right side of this curve, the time average damping is positive. To the left of this curve in the $(\delta,\epsilon)$ parameter space, the averaged damping is negative, and this leads to characteristic multipliers $\lambda(M)>1$. Even if the $\overline{\xi} > 0$, but $\xi(t)<0$ for some $t \in [0,T]$, then it is possible for $\lambda(M)>1$ as shown by the small unstable parametric region to the left of the black curve in fig.(\ref{fig:stability_A}). The parametric region in red (indicated by \textbf{U}) on the left of the black line has average damping less than zero ($\overline{\xi} < 0$), and the region on the right (indicated by \textbf{S})  has positive average damping ($\overline{\xi} >0$). 
 Also, with the increase in the value of $C_{\psi}$, the damping term increases, which increases the stable region as shown in fig.(\ref{fig:stability_A}b). The figure also shows that the linear system is unstable for the negative values of the average damping as it propagates motion in the roll direction \cite{Batchelor_damp}. Interestingly we observe a small unstable region on the right side of the $\overline{\xi}=0$ curve. This region is part of the parametric resonance instability around $\delta = 0.25$ and $\delta = 1$.  The unstable region due to average negative damping happens to be overlaid on the usual unstable region due to parametric resonance.

To investigate the unstable region due to parametric resonance, we choose to consider parameters such that the average damping remains positive. To construct the stability graph we first find out the parameter values $[A,~\Omega,~u_c,~\alpha_s,~\alpha_c,~\beta_s,~\beta_c, ~C_{\psi}]$ for each point in $\delta-\epsilon$ space using the equations \eqref{eq:parameter_eqn}(a-e), \eqref{eq:delta} and \eqref{eq:epsilon} along with average damping equation \eqref{eq:avg_damping}
and then perform Floquet analysis.
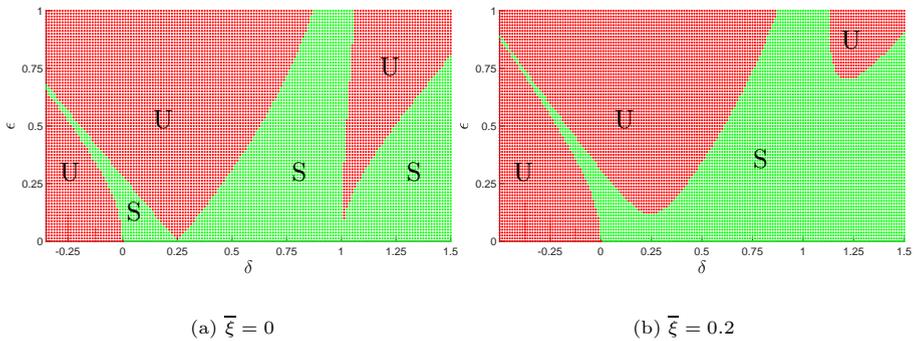
\begin{figure}[h]
         \begin{picture}(360,100)\put(0,0){\includegraphics[width=0.5\columnwidth]{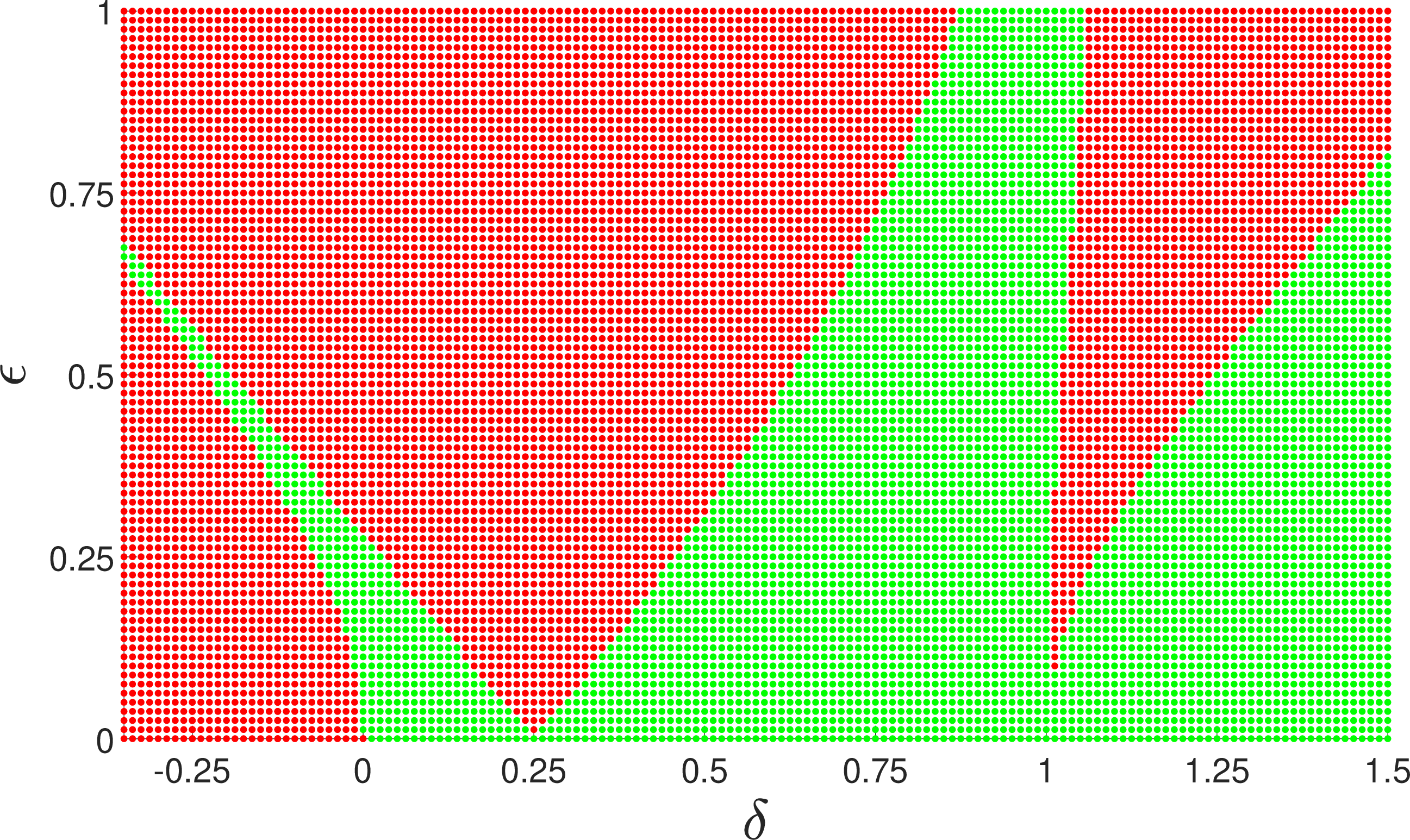}}\put(170,0){\includegraphics[width=0.5\columnwidth]{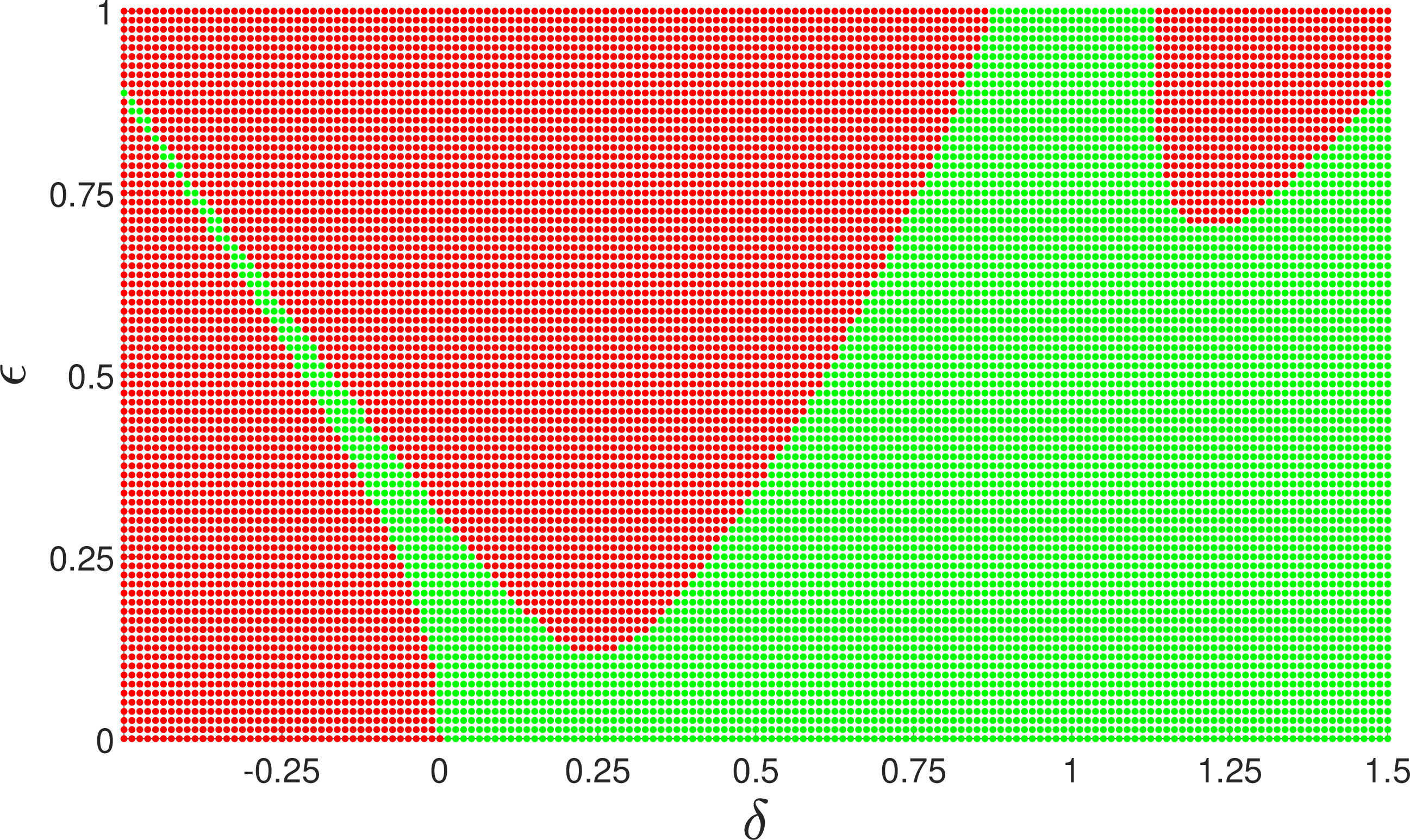}}\put(190,35){U}\put(20,35){U}\put(45,20){S} \put(55,55){U} \put(228,55){U} \put(107,35){S} \put(280,40){S} \put(140,75){U} \put(150,35){S} \put(313,85){U}\end{picture}
         \subfloat[$\overline{\xi} = 0 $]{\hspace{.5\linewidth}}
         \subfloat[$\overline{\xi} = 0.2$]{\hspace{.5\linewidth}}
        \caption{Keeping average of damping term $\xi$ constant we observe stability chart for the equation \eqref{eq:mathieu} in $\delta - \epsilon$ parameter space is numerically obtained using Floquet theory.}\label{fig:stability_B}
\end{figure}
The fig.(\ref{fig:stability_B}) shows the stability chart for the linear homogeneous parametric oscillator \eqref{eq:hom_lin} for constant average damping. The stability chart in fig.(\ref{fig:stability_B}) resembles the Mathieu stability chart with some variations in the size of the instability tongue, which occurs due to periodic damping. We observe that the system near marginal stability is susceptible to the phenomenon in which instability may be caused due to modulation of the damping coefficient, which is negative for a fraction of the cycle. This phenomenon is also an example of parametric resonance \cite{Batchelor_damp}. The system instability cannot be dynamically stabilized by periodically modulating the damping term $\xi(\tau)$ because the constant term is much larger than the periodic term in $\xi(\tau)$. 

The damping in the homogeneous system \eqref{eq:hom_lin} effectively reduces the unstable region as shown in figs.(\ref{fig:mathieu_chart},\ref{fig:stability_A} and \ref{fig:stability_B}) and through numerical simulation it could also be shown that damping reduces the magnitude of response. All of the simulations used the following numerical values; we assume that the sleigh has a uniform density of $990$ Kg/m$^{3}$, which is slightly lighter than water, and also it has the shape of a prolate spheroid with a ratio of major and minor axis equal to $2.5$. We further assume $b=0.125$ m, $h=0.05$ m, $C_u=0.5$ kg/s and $C_{\theta} = 0.3$ kg/s.


\subsection{Linearized non-homogeneous parametric oscillator} \label{sec:lin_nonhom}
\begin{figure}[htbp]
\subfloat[Linear Homogeneous system \eqref{eq:hom_lin}]
{\centering
\includegraphics[width=0.5\columnwidth]{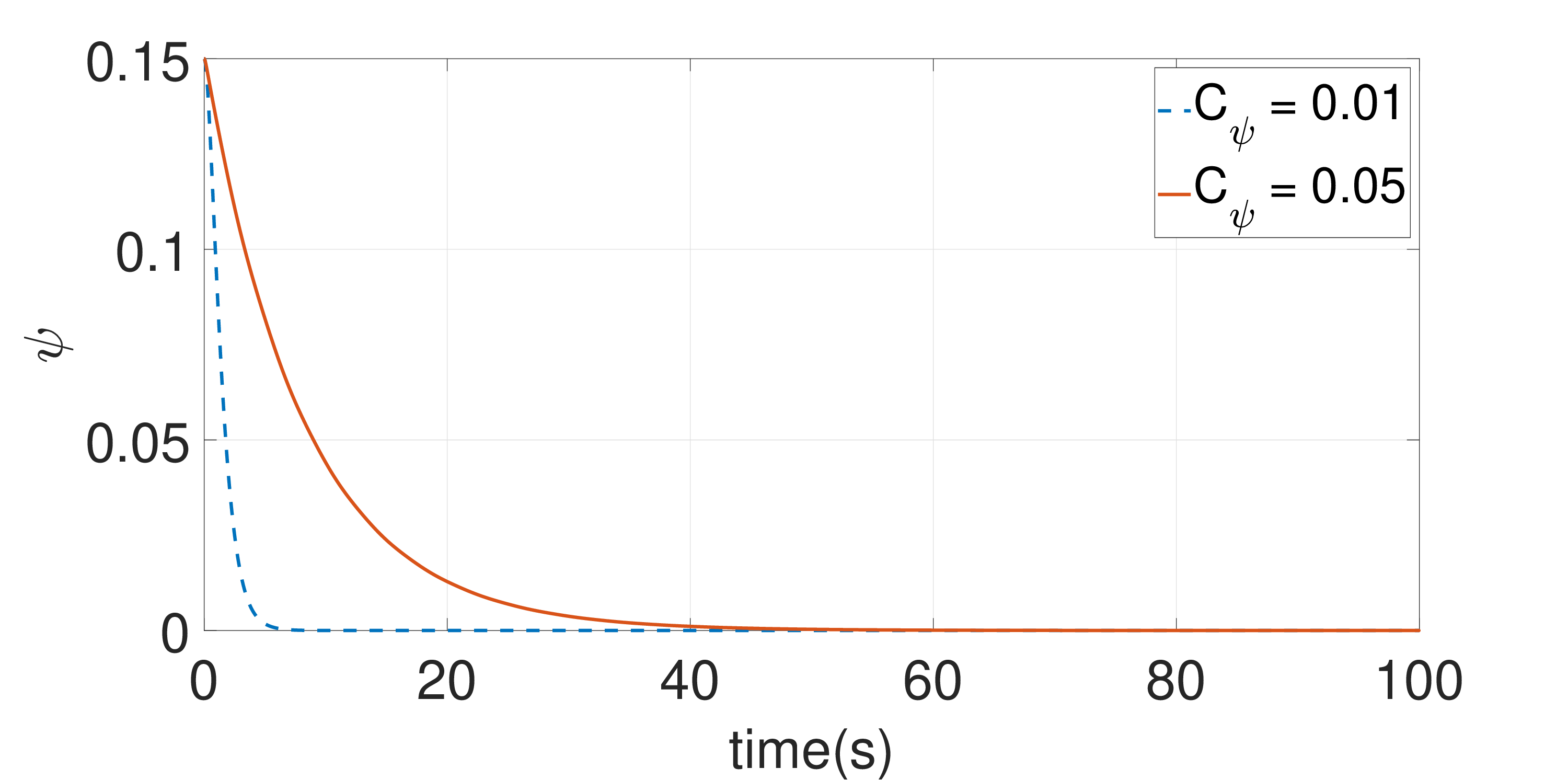}
}
\subfloat[Linear Non-Homogeneous system (\eqref{eq:non_hom_lin})]
{\centering
\includegraphics[width=0.5\columnwidth]{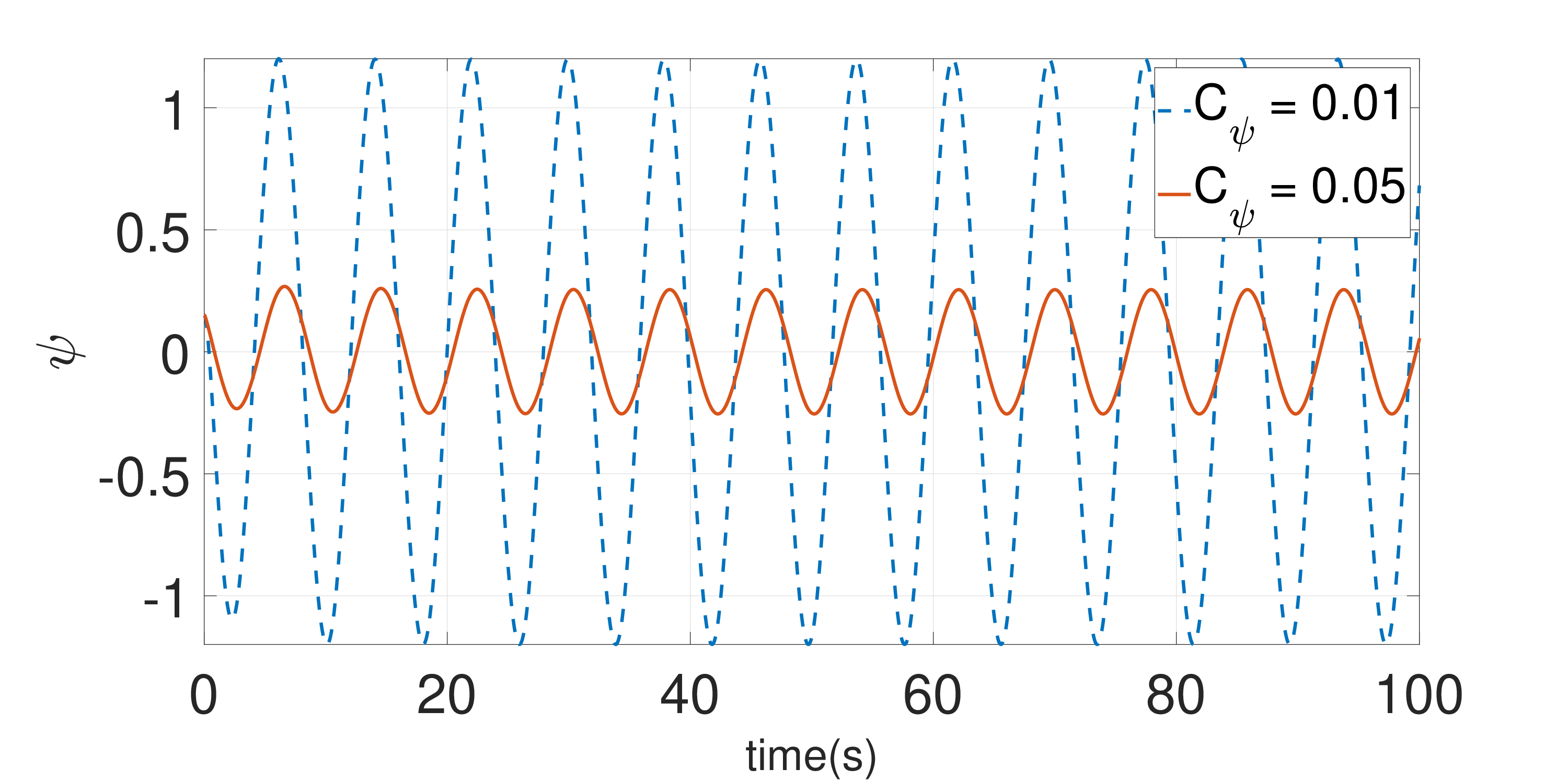}
}
\caption{Part (a) shows the response of linear homogeneous system \eqref{eq:hom_lin} and part (b) shows the response of Linear non-homogeneous system \eqref{eq:non_hom_lin} with parameters $\delta = 0.7$ and $\epsilon = 0.02$ with varying the $C_{\psi}$ value. For dotted blue trajectory $C_{\psi} = 0.01$ and for solid red line $C_{\psi} = 0.05$} \label{fig:hom_linear} 
\end{figure}

The effect of non-homogeneous term $C(\tau)$ on the solution of Hill's equation has been studied in the past, for instance by \cite{YOUNESIAN200758,SHADMAN200568,Slane2011AnalysisOP,Rodriguez2015OnSO}. Slane and Tragesser in \cite{Slane2011AnalysisOP} modified Floquet theory to use Floquet multipliers for analytically examining transitory and steady-state behavior of the non-homogeneous system, in which parametric and forcing excitations have the same time period ($C(\tau + T) = C(\tau)$). The introduction of forced excitation changes the fundamental behavior of the homogeneous system in only two ways. First, when $\lvert \lambda(\mathbf{M}) \rvert < 1$ the solution of the non-homogenous Hills equations changes to bounded from asymptotic stability. The numerical simulation of the linear system without the forcing \eqref{eq:hom_lin}  and with forcing \eqref{eq:non_hom_lin} shows the anticipated solution where an asymptotic stable solution in fig.(\ref{fig:hom_linear}a) changes to a bounded solution fig.(\ref{fig:hom_linear}b).

 The second way in which the forced excitation affects the response is when the unit characteristic multipliers are simple roots of the minimal polynomial of the Monodromy matrix. Under this condition, the solution of the system changes from Lyapunov stable to unbounded. Linear resonance also occurs if the natural frequency of the roll oscillation is equal to the frequency of the direct forcing term $C(\tau)$. For  \eqref{eq:non_hom_lin} we have non-homogeneous term \eqref{eq:nonhom} which we can represent compactly as 
\begin{equation}
     C(\tau) =\sum_{\mathbf{k}=1\&3} \nu_{\mathbf{c},k} \cdot \cos (k\cdot \frac{\tau }{2})+\nu_{\mathbf{s},k} \cdot \sin (k\cdot \frac{\tau }{2})
\end{equation}
 where ($\nu_{\mathbf{c},k},~\nu_{\mathbf{s},k}$) are constants dependent on limit-cycle parameters. The direct forcing term has time period $2T$ and $\frac{2T}{3}$ where $T = 2\pi$. It is known through corollary to \textit{Floquet-Lyapunov} theorem in \cite{Magnus_Winkler_Hillseqn}, stable regions in fig.(\ref{fig:stability_B}) of the homogeneous system \eqref{eq:hom_lin}  has non-trivial $kT$-periodic solutions $ k\in \mathbf{N}>2$; but the solutions corresponding to period $2T$ and $\frac{2T}{3}$ are already in the unstable region and hence we do not observe linear resonance.

\begin{figure}[htbp]
\centering\subfloat[$\overline{\xi} = 0$]
{\centering\includegraphics[width=0.5\columnwidth]{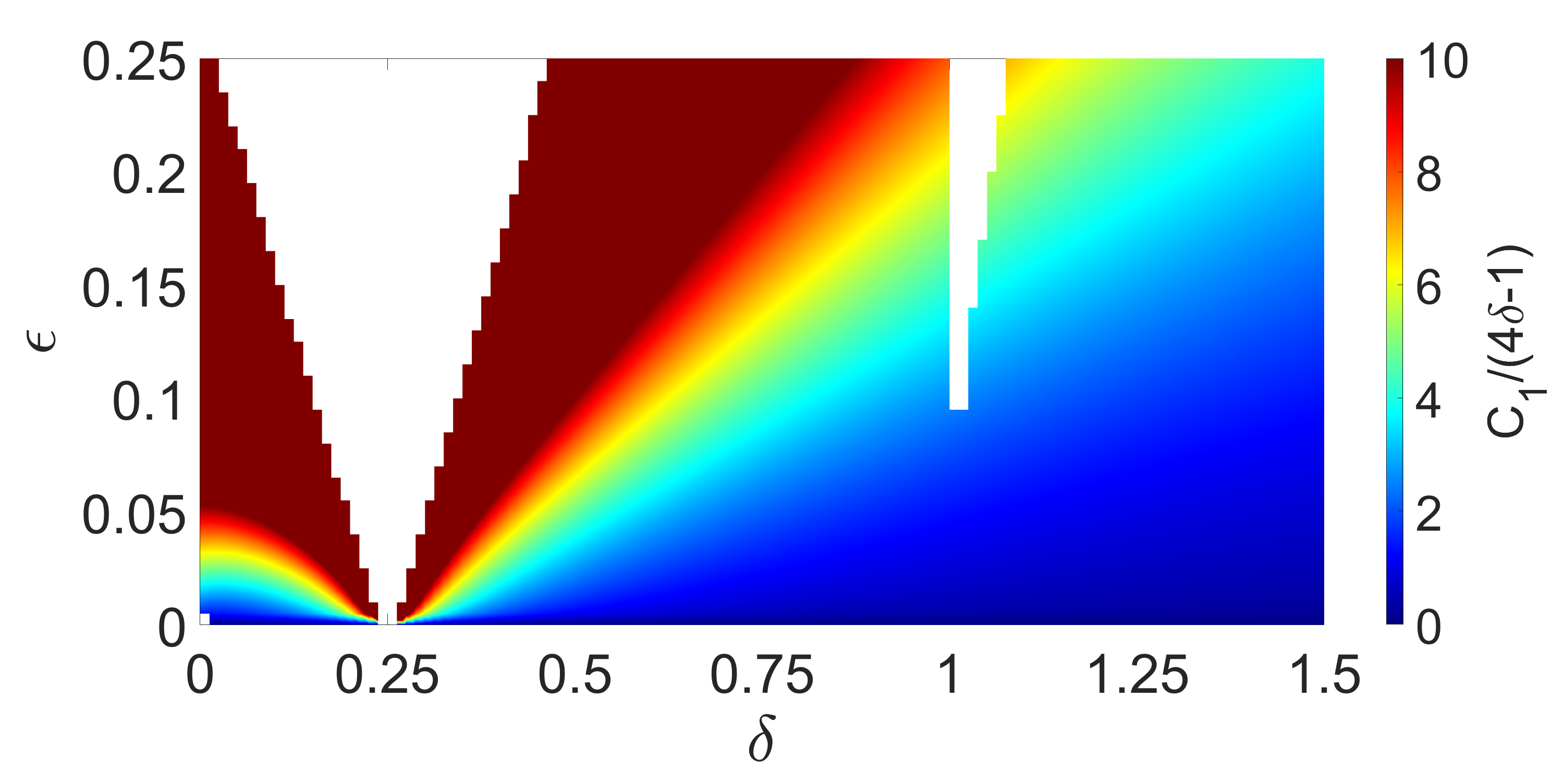}} \subfloat[$\overline{\xi} = 0.2$]
{\centering\includegraphics[width=0.5\columnwidth]{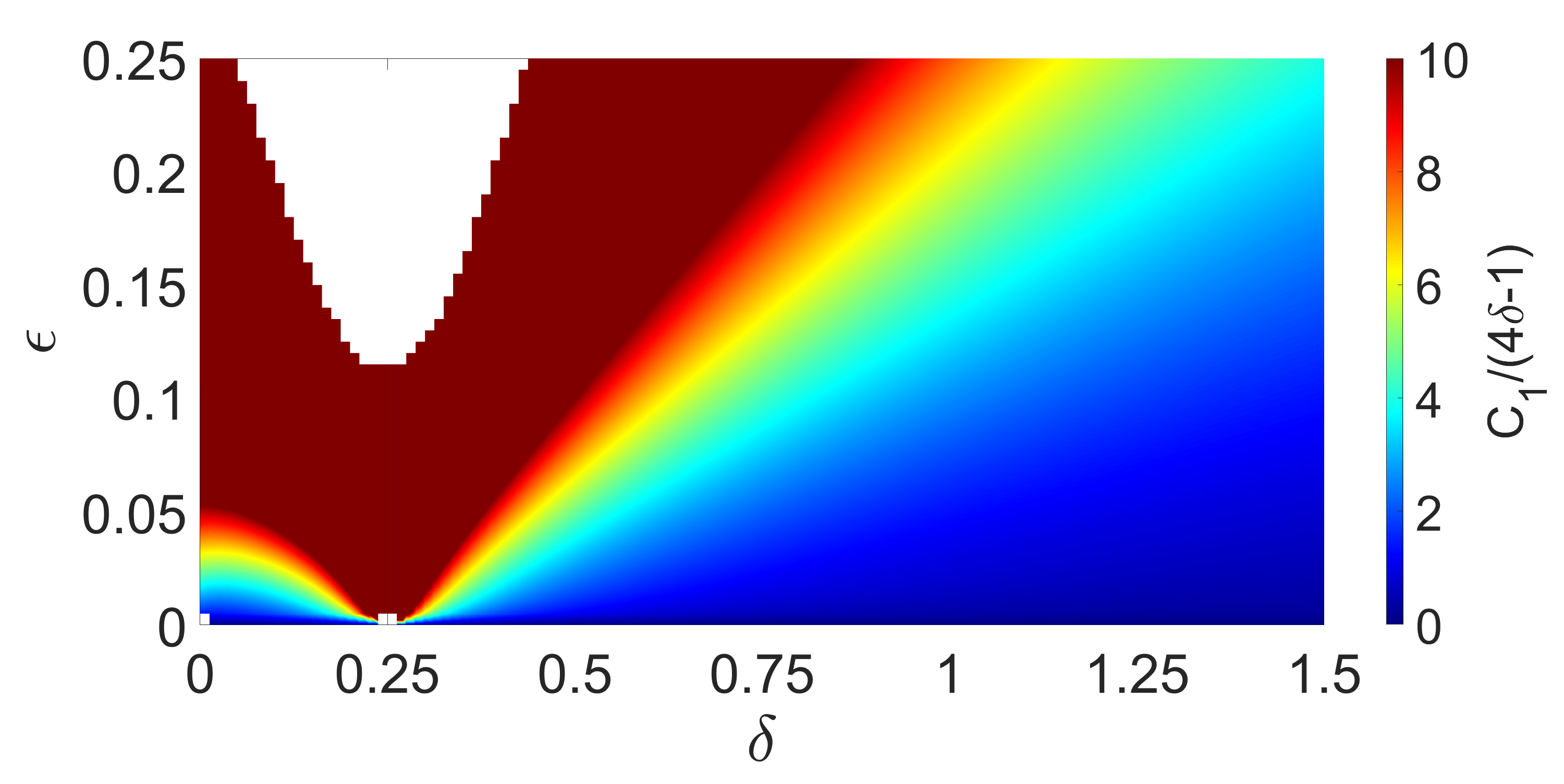}}
\caption{A magnitude plot of $\frac{C_1}{4\delta-1}$ over a range of $\epsilon$ and $\delta$ for (a) $\overline{\xi} =0$ and (b) $\overline{\xi} =0.2$. Parametric regions where $\frac{C_1}{4\delta-1}$ becomes unbounded are shown in white. } \label{fig:Coeff_range} \end{figure}

All bounded solutions are not necessarily desirable solutions for the physical system described by \eqref{eq:non_hom_lin}.  Large amplitude roll oscillations can also be produced due to the coefficients ($\nu_{c,i}, \nu_{s,i}$) of the direct forcing term. To see the effect of forcing term on the linear parametric oscillator system \eqref{eq:non_hom_lin}, an analysis using multiple scale method \cite{Mathieu_RRand,nayfeh2008perturbation} is presented. The stretched time scale will be denoted by $t_0 = \tau$ and the slow time scale $t_1 = \epsilon \tau$. The solution for $\psi$ can be expanded into a power series as 
\begin{equation}\label{eq:power_series}
    \psi = \psi_0(t_0,t_1) + \epsilon \psi_1(t_0,t_1) + \epsilon^2 \psi_2(t_0,t_1) + \mathcal{O}(3)
\end{equation}
For the system \eqref{eq:non_hom_lin} let's consider the case where $\overline{\xi} = 0$. Then we can rewrite the system \eqref{eq:non_hom_lin} as 
\begin{equation}\label{eq:nonhom_lin_msm}
    \ddot{\psi} + \Big(\delta + \epsilon \cos(\tau-\gamma)\Big)\psi +\Big(\alpha_c'\cos\tau + \alpha_s'\sin\tau\Big)\dot{\psi} =  C_1 \cos (\frac{\tau}{2}-\nu_1)+C_3 \cos (\frac{3\tau }{2} - \nu_3)
\end{equation}
Where, $\alpha_s' = \xi_2 \alpha_s$ and $\alpha_c' = \xi_2 \alpha_c$ are of the order $\epsilon$. We choose the following transformations to compactly represent the equation \eqref{eq:nonhom_lin_msm}.
\begin{align*}
    C_{1} =& \sqrt{\nu_{c,1}^2 + \nu_{s,1}^2}& \nu_1 = \arccos{\Bigg(\frac{\nu_{c,1}}{\sqrt{\nu_{c,1}^2 + \nu_{s,1}^2}}\Bigg)}\\
    C_{3} =& \sqrt{\nu_{c,3}^2 + \nu_{s,3}^2}&\nu_3 = \arccos{\Bigg(\frac{\nu_{c,3}}{\sqrt{\nu_{c,1}^2 + \nu_{s,3}^2}}\Bigg)}
\end{align*}
A direct substitution of Eq.\eqref{eq:power_series} into Eq.\eqref{eq:nonhom_lin_msm} and setting $\frac{d}{d\tau} = \frac{\partial}{\partial t_0} + \epsilon \frac{\partial}{\partial t_1}$. Then equating terms of $\mathcal{O}(\epsilon^0)$ and $\mathcal{O}(\epsilon^1)$ yield the following equations,
\begin{equation}\label{eq:order_e0}
    \frac{\partial^2 \psi_0}{\partial t_0^2} + \delta\psi_0 = C_1 \cos\Big(\frac{t_0}{2}-\nu_1\Big) + C_3 \cos\Big(\frac{3t_0}{2}-\nu_3\Big)
\end{equation}
\begin{equation}\label{eq:order_e1}
    \frac{\partial^2 \psi_1}{\partial t_0^2} + \delta\psi_1 = -2\frac{\partial^2\psi_0}{\partial t_0 \partial t_1} - \cos(t_0-\gamma)\psi_0 - \Big(\alpha_c'\cos\tau + \alpha_s'\sin\tau\Big)\frac{\partial \psi_0}{\partial t_0}
\end{equation}
The solution to Eq.\eqref{eq:order_e0} is 
\begin{align}\label{eq:sol_order_e0}
    \begin{split}
        \psi_0(t_0,t_1) = &A(t_1)\cos\sqrt{\delta}t_0 + B(t_1)\sin\sqrt{\delta}t_0 + \frac{4C_1}{4\delta -1}\cos\Big(\frac{t_0}{2}-\nu_1\Big)\\ & + \frac{4C_3}{4\delta -9}\cos\Big(\frac{3t_0}{2}-\nu_3\Big)
    \end{split}
\end{align}
which when substituted in Eq.\eqref{eq:order_e1} we get,
\begin{align}\label{eq:order_e10}
    \frac{\partial^2 \psi_1}{\partial t_0^2} &+ \delta \psi_1 = \frac{\partial A}{\partial t_1}\sqrt{\delta} \sin \sqrt{\delta}t_0 - \frac{\partial B}{\partial t_1}\sqrt{\delta}  \cos\sqrt{\delta}t_0  - \nonumber\\ & \frac{\cos(\sqrt{\delta}-1)t_0}{2}\Big(A\cos\gamma + B\sin\gamma + \alpha_c'B\sqrt{\delta} - \alpha_s'A\sqrt{\delta} \Big) - \nonumber\\ & \frac{\cos(\sqrt{\delta}+1)t_0}{2}\Big(A\cos\gamma - B\sin\gamma + \alpha_c'B\sqrt{\delta} + \alpha_s'A\sqrt{\delta} \Big) + \nonumber\\ & \frac{\sin(\sqrt{\delta}-1)t_0}{2}\Big(A\cos\gamma - B\sin\gamma + \alpha_c'B\sqrt{\delta} + \alpha_s'A\sqrt{\delta} \Big) - \nonumber\\ & \frac{\sin(\sqrt{\delta}+1)t_0}{2}\Big(A\cos\gamma + B\sin\gamma - \alpha_c'B\sqrt{\delta} + \alpha_s'A\sqrt{\delta} \Big) + \nonumber\\ & \frac{C_1}{4\delta - 1}\Bigg(\cos\frac{t_0}{2}\Big(-2\cos(\gamma-\nu_1)-\alpha_c'\sin\nu_1 + \alpha_s'\cos\nu_1\Big) + \nonumber\\&~~~~~ \sin\frac{t_0}{2}\Big(2\sin(\gamma-\nu_1)-\alpha_c'\cos\nu_1 - \alpha_s'\sin\nu_1\Big)\Bigg) +\\& \frac{C_2}{4\delta-9}\Bigg(\cos\frac{t_0}{2}\Big(2\cos(\nu_3-\gamma)+3\alpha_c'\sin\nu_3 - 3\alpha_s'\cos\nu_3\Big) +  \nonumber\\ &~~~~~ \sin\frac{t_0}{2}\Big(2\sin(\nu_3-\gamma)+3\alpha_c'\cos\nu_3 + 3\alpha_s'\sin\nu_3\Big)\Bigg) + \nonumber\\  & \frac{C_1 }{4\delta-1}\Bigg(\cos\frac{3t_0}{2}\Big(-2\cos(\nu_1+\gamma) - \alpha_c'\sin\nu_1 -\alpha_s'\cos\nu_1 \Big) +\nonumber \\ & ~~~~~\sin\frac{3t_0}{2}\Big(2\sin(\nu_1+\gamma) + \alpha_c'\cos\nu_1 -\alpha_s'\sin\nu_1 \Big)\Bigg)  \nonumber\\ & \frac{C_2 }{4\delta-9}\Bigg(\cos\frac{5t_0}{2}\Big(-2\cos(\nu_3+\gamma) - 3\alpha_c'\sin\nu_3 -3\alpha_s'\cos\nu_3 \Big) +\nonumber \\ & ~~~~~\sin\frac{5t_0}{2}\Big(2\sin(\nu_3+\gamma) + 3\alpha_c'\cos\nu_3 -3\alpha_s'\sin\nu_3 \Big)\Bigg)  \nonumber
\end{align}
We observe from the above eq.\eqref{eq:order_e10} that the resonant terms occur for four possible values of non-dimensional frequency $\delta$. The first, when $\delta = 1$ and the resonant terms are $\cos\sqrt{\delta}t_0$ and $\sin\sqrt{\delta}t_0$. The second case occurs when $\delta = \frac{1}{4}$ and the resonant terms are $\cos(\sqrt{\delta}-1)t_0,~ \sin(\sqrt{\delta}-1)t_0, ~\cos\frac{t_0}{2} $ and $\sin\frac{t_0}{2}$. Also for $\delta = 9/4$ and $\delta = 25/4$ we have resonant terms $[\cos\frac{3t_0}{2}, \sin\frac{3t_0}{2}]$ and $[\cos\frac{5t_0}{2}, \sin\frac{5t_0}{2}]$ respectively.   The unstable regions in the parameter space for the system \eqref{eq:nonhom_lin_msm} emanate with their vertex at $(\delta = \frac{n^2}{4}, \epsilon = 0)$ for $n>0$; these critical values of $(\delta, \epsilon)$ around which the solutions to $\psi$ diverge are same as those of the homogeneous Mathieu equation \cite{Mathieu_RRand}. However, interestingly the resonances occurring at $\delta = 9/4$ and $\delta = 25/4$ show up in the $O(\epsilon)$ equation \eqref{eq:order_e10} as opposed to higher order equation. Even more importantly, the coefficient terms such as $\frac{C_1}{4\delta-1}$ and $\frac{C_2}{4\delta -9}$ will diverge at $\delta = \frac{1}{4}$ and $\delta = \frac{9}{4}$ respectively and have large values outside the parametric resonant unstable regions. Figure (\ref{fig:Coeff_range}) shows the range of values of the coefficient $\frac{C_1}{4\delta-1}$ over the $\delta -\epsilon$ space; it shows that the forcing coefficient is very high even outside but adjacent to the resonance tongue of instability. This causes large amplitude roll oscillations even outside the unstable resonant parametric regions of the Mathieu equation.

\section{Numerical simulations of the nonlinear equations} \label{sec:numerics}
\begin{figure}[htbp]\centering
\includegraphics[width=0.75\columnwidth]{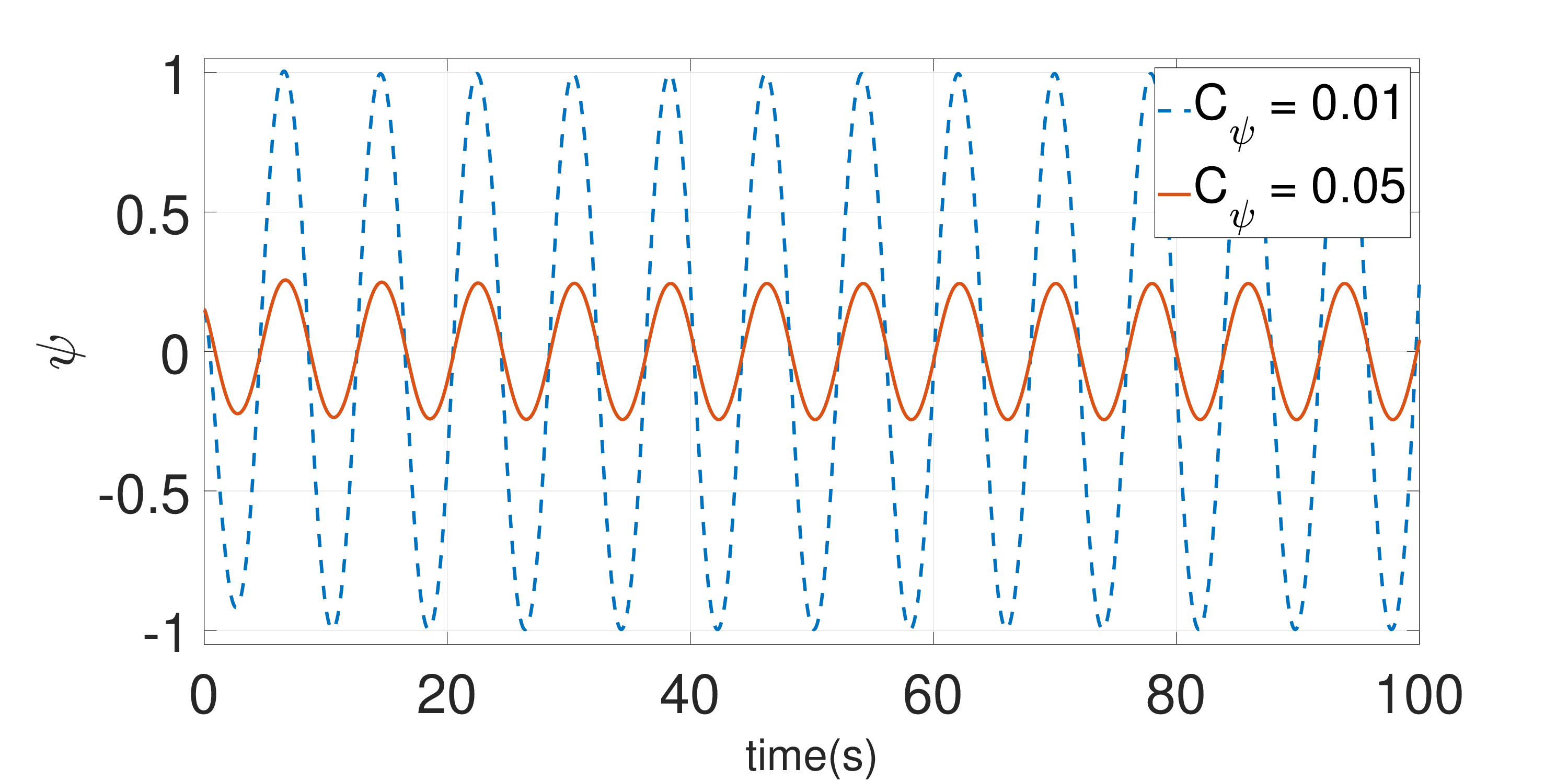}\caption{Shows response of non-linear hydrodynamic Chaplygin sleigh with roll dynamics \eqref{eq:full} simulated with similar parameters in fig.(\ref{fig:hom_linear}) that is $\delta = 0.7,~ \epsilon = 0.02$. Again for dotted blue trajectory $C_{\psi} = 0.01$ and for red solid line $C_{\psi} = 0.05$.} \label{fig:non_linear} 
\end{figure}

The analysis of the reduced linear parametric oscillator \eqref{eq:non_hom_lin} in section \ref{sec:linear} first relies on the reduction of the dimension of the dynamical system \eqref{eq:full} to a reduced one degree of freedom oscillator \eqref{eq:psidot} and then linearizing this equation about its fixed point. The solutions of $\psi(t)$ from such analysis are compared to the numerical solution of the roll oscillations of the unreduced non-linear hydrodynamic Chaplygin sleigh \eqref{eq:full}.  To demonstrate the validity of the linear analysis, we choose parameters $(\delta = 0.7, \epsilon = 0.02)$ and $C_{\psi} = 0.01$ (dotted blue line) and $C_{\psi} = 0.05$ (solid red line) and plot the roll angle $\psi(t)$ from a simulation of equation \eqref{eq:full} in fig. \ref{fig:non_linear}. Comparing the solutions in fig. \ref{fig:non_linear} with the solution of the reduced parametric nonhomogeneous oscillator in fig. \ref{fig:hom_linear}(b), it can be seen that the frequency and amplitude of the two solutions are nearly indistinguishable. 

\begin{figure}[htbp]\centering\subfloat[$\overline{\xi} = 0$]{\centering\includegraphics[width=0.5\columnwidth]{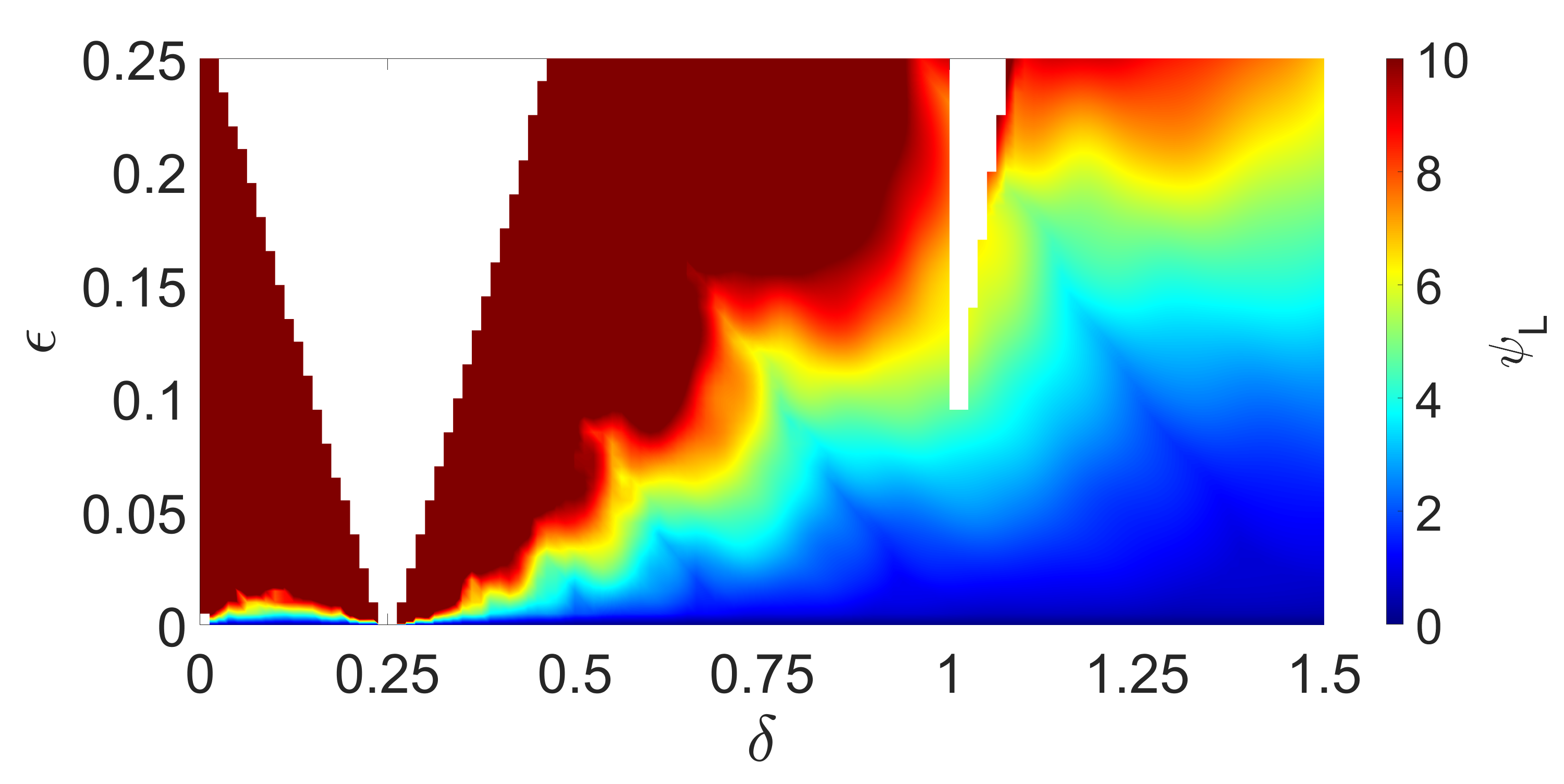}} \subfloat[$\overline{\xi} = 0.2$]{\centering\includegraphics[width=0.5\columnwidth]{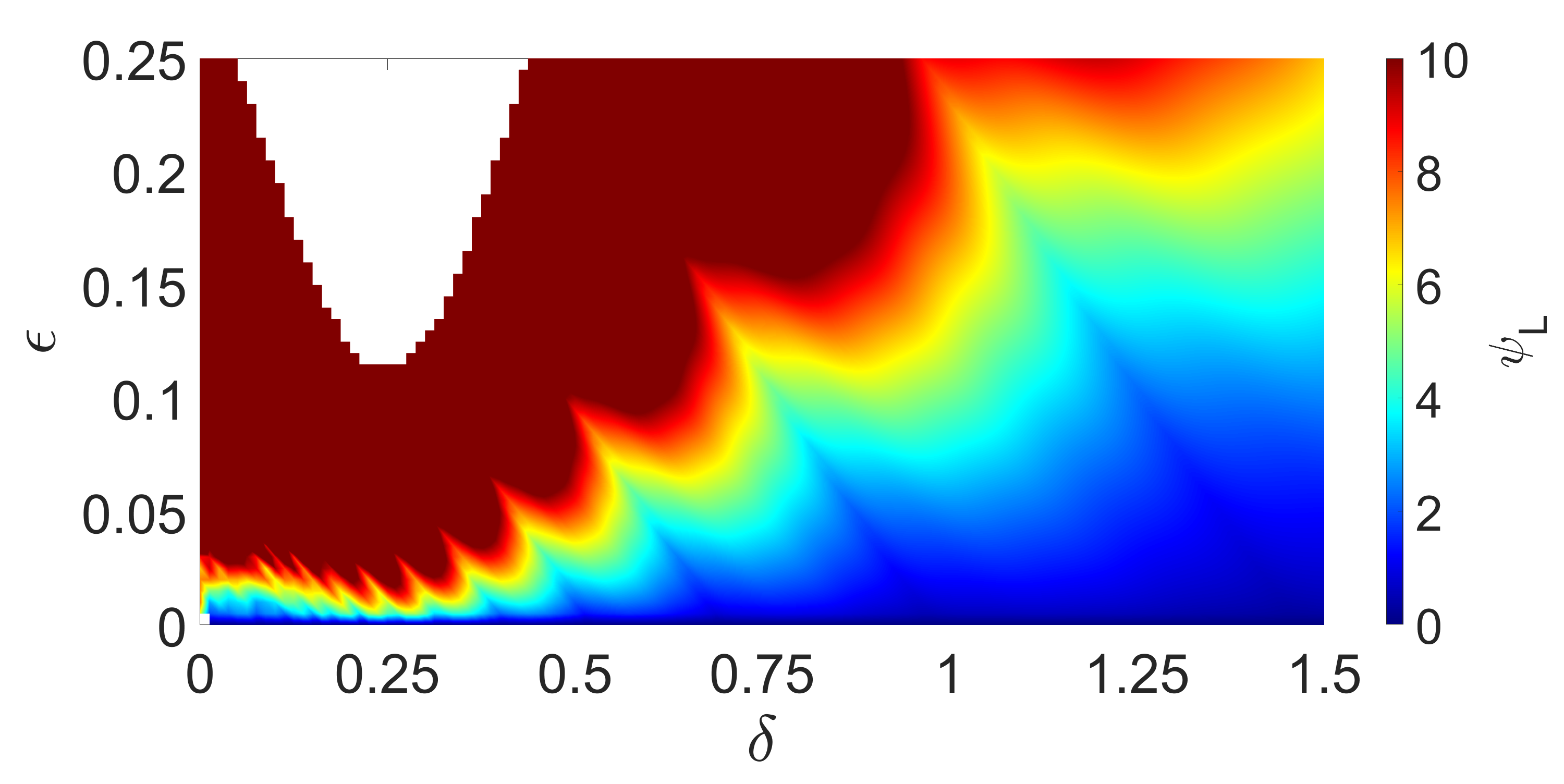}}\caption{A plot for the amplitude of $\psi$ over a range of $\delta$ and $\epsilon$ obtained from the solution of linear non-homogeneous \eqref{eq:mathieu} system for (a) $\overline{\xi} =0$ and (b) $\overline{\xi} =0.2$. Parametric regions where amplitudes of $\psi$ are large enough to make the system unstable due to the term $\frac{C_1}{4 \delta - 1}$ (refer fig.\ref{fig:Coeff_range}).}\label{fig:error_amp} \end{figure}

The parameters $(\delta,\epsilon)$ for this comparison were chosen so as to lie in a region where the coefficient $\frac{C_1}{4\delta-1} \approx 1$ (see fig. \ref{fig:Coeff_range}). If $(\delta,\epsilon)$ is chosen such that $\frac{C_1}{4\delta-1}$ is large, the amplitude of the linear solution becomes large, rendering the linearization a poor approximation. Figure \ref{fig:error_amp} shows the amplitude of the roll oscillations of the linear non-homogeneous system \eqref{eq:non_hom_lin} obtained through a direct numerical simulation. We can observe looking at fig. \ref{fig:error_amp} and \ref{fig:Coeff_range} the amplitudes of $\psi$ are high where the magnitude of the forcing in \eqref{eq:non_hom_lin} is high.

\section{Conclusion}\label{sec:conclusion}
The hydrodynamic Chaplygin sleigh model considered in this paper is motivated by applications to robotic locomotion at the bottom of a body of water and fish-like swimming robots where the propulsion is made possible through periodic torques. For simplicity, we considered the motion of a single rigid body as opposed to an articulated body. The analysis of this four degrees of freedom system is simplified first by the presence of a nonholonomic constraint and then further by observing the existence of limit cycles in the velocity space of a planar Chaplygin sleigh. When the four-degree of freedom Chaplygin sleigh is subject to a periodic torque, its longitudinal velocity and yaw angular velocity closely track the limit cycles of the planar Chaplygin sleigh. This observation leads to the approximation where the planar motion is independent of the roll dynamics, but the roll dynamics are coupled with the planar motion. Linearization of the roll equation then leads to a parametric oscillator. Stability analysis of these linear homogeneous systems using Floquet theory leads to a stability chart in two parameter spaces. When the hydrodynamic added mass tensor is purely diagonal, and all the diagonal terms are equal (such as for a sphere), the roll dynamics are modeled by a Mathieu equation, and the stability chart is identical to that of the Mathieu oscillator. A different stability chart than the Mathieu stability chart is obtained for the case where added mass terms are not the same. These charts show that the condition for instability is always satisfied if the time average damping is negative and parametric resonance can exist for positive average damping. This analysis further showed that higher average longitudinal velocity ($u_c$) reduces the effects of damping and increases the amplitude of roll direction.

The trade-off between stability, maneuverability, and speed in the locomotion of fish is a   phenomenon that is observed in several species of fish \cite{Webb2001}. A fish or a robot shaped as slender body usually can be a faster swimmer but at the same time, this geometry renders it susceptible to roll instability due to perturbations arising from fish-like body-caudal fin (BCF) propulsion can cause instability in the roll motion. The design of various biomimetic fish robots is often faced with the trade-off between stability on the one hand and speed and maneuverability on the other.  While such trade-offs and the effect of the shape on roll stability have been observed in different species of fish and can be expected to occur in both free-swimming robots and robots that move on the bottom of a body of water, a systematic analysis from the perspective of nonlinear dynamics has been absent.  The analysis of the parametric roll dynamics of the hydrodynamic Chaplygin sleigh in this paper is the first investigation of such phenomena. The Chaplygin sleigh considered in this paper has much of the physics applicable for a body crawling with a nonholonomic constraint on the bottom of a pool and has similarities to models of fish-like motion. While the model does not necessarily simulate the fluid-body interaction fully, it captures the inviscid effects imposed by added mass and the Kutta-Joukowski condition, and the results in this paper have utility for the design of swimming robots. Design considerations such as shape, the inertia tensor, added mass tensor, and choice of periodic excitation and gaits can be informed by the framework of the analysis in this paper. Further work motivated by this paper can extend the results to the case of articulated slender bodies with elastic joints and coupling of the roll, pitch, and yaw dynamics with rigid body motion and provide a more detailed picture of the stability and agility of a fish-like swimmer.

\section*{Statements and Declarations }
\subsection*{Funding}
This work was partially supported by the NSF Grant 2021612 and ONR Grant 13204704.

\subsection*{Competing Interests}
The authors have no relevant financial or non-financial interests to disclose.

\subsection*{Author Contributions}
All authors contributed to the study conception and design. All authors read and approved the final manuscript.

\section*{Data Availability Statement}
The code that was used for simulations and analysis will be available from the corresponding author on reasonable request. 




\bibliography{fish3.bib}


\begin{thebibliography}{49}
\ifx \bisbn   \undefined \def \bisbn  #1{ISBN #1}\fi
\ifx \binits  \undefined \def \binits#1{#1}\fi
\ifx \bauthor  \undefined \def \bauthor#1{#1}\fi
\ifx \batitle  \undefined \def \batitle#1{#1}\fi
\ifx \bjtitle  \undefined \def \bjtitle#1{#1}\fi
\ifx \bvolume  \undefined \def \bvolume#1{\textbf{#1}}\fi
\ifx \byear  \undefined \def \byear#1{#1}\fi
\ifx \bissue  \undefined \def \bissue#1{#1}\fi
\ifx \bfpage  \undefined \def \bfpage#1{#1}\fi
\ifx \blpage  \undefined \def \blpage #1{#1}\fi
\ifx \burl  \undefined \def \burl#1{\textsf{#1}}\fi
\ifx \doiurl  \undefined \def \doiurl#1{\url{https://doi.org/#1}}\fi
\ifx \betal  \undefined \def \betal{\textit{et al.}}\fi
\ifx \binstitute  \undefined \def \binstitute#1{#1}\fi
\ifx \binstitutionaled  \undefined \def \binstitutionaled#1{#1}\fi
\ifx \bctitle  \undefined \def \bctitle#1{#1}\fi
\ifx \beditor  \undefined \def \beditor#1{#1}\fi
\ifx \bpublisher  \undefined \def \bpublisher#1{#1}\fi
\ifx \bbtitle  \undefined \def \bbtitle#1{#1}\fi
\ifx \bedition  \undefined \def \bedition#1{#1}\fi
\ifx \bseriesno  \undefined \def \bseriesno#1{#1}\fi
\ifx \blocation  \undefined \def \blocation#1{#1}\fi
\ifx \bsertitle  \undefined \def \bsertitle#1{#1}\fi
\ifx \bsnm \undefined \def \bsnm#1{#1}\fi
\ifx \bsuffix \undefined \def \bsuffix#1{#1}\fi
\ifx \bparticle \undefined \def \bparticle#1{#1}\fi
\ifx \barticle \undefined \def \barticle#1{#1}\fi
\bibcommenthead
\ifx \bconfdate \undefined \def \bconfdate #1{#1}\fi
\ifx \botherref \undefined \def \botherref #1{#1}\fi
\ifx \url \undefined \def \url#1{\textsf{#1}}\fi
\ifx \bchapter \undefined \def \bchapter#1{#1}\fi
\ifx \bbook \undefined \def \bbook#1{#1}\fi
\ifx \bcomment \undefined \def \bcomment#1{#1}\fi
\ifx \oauthor \undefined \def \oauthor#1{#1}\fi
\ifx \citeauthoryear \undefined \def \citeauthoryear#1{#1}\fi
\ifx \endbibitem  \undefined \def \endbibitem {}\fi
\ifx \bconflocation  \undefined \def \bconflocation#1{#1}\fi
\ifx \arxivurl  \undefined \def \arxivurl#1{\textsf{#1}}\fi
\csname PreBibitemsHook\endcsname

\bibitem{bloch03}
\begin{botherref}
\oauthor{\bsnm{Bloch}, \binits{A.M.}}:
Nonholonomic mechanics and control.
Springer
(2003)
\end{botherref}
\endbibitem

\bibitem{V.Borisov2007}
\begin{botherref}
\oauthor{\bsnm{V.Borisov}, \binits{A.}},
\oauthor{\bsnm{Mamaev}, \binits{I.S.}},
\oauthor{\bsnm{Ramodanov}, \binits{S.M.}}:
Dynamic interaction of point vortices and a two-dimensional cylinder.
Journal of Mathematical Physics
(2007)
\end{botherref}
\endbibitem

\bibitem{Neimark1972}
\begin{botherref}
\oauthor{\bsnm{Neimark}, \binits{J.I.}},
\oauthor{\bsnm{Fufaev}, \binits{N.A.}}:
Dynamics of nonholonomic systems.
AMS
(1972)
\end{botherref}
\endbibitem

\bibitem{lauder_afm_2015}
\begin{barticle}
\bauthor{\bsnm{Lauder}, \binits{G.V.}}:
\batitle{Fish locomotion: recent advances and new directions}.
\bjtitle{Annual review of marine science}
\bvolume{7},
\bfpage{521}--\blpage{545}
(\byear{2015})
\end{barticle}
\endbibitem

\bibitem{triantafyllou_afm_2016}
\begin{botherref}
\oauthor{\bsnm{Triantafyllou}, \binits{M.S.}},
\oauthor{\bsnm{Weymouth}, \binits{G.D.}},
\oauthor{\bsnm{Miao}, \binits{J.}}:
Biomimetic survival hydrodynamics and flow sensing.
Annual Review of Fluid Mechanics
\textbf{48}(1)
(2016)
\end{botherref}
\endbibitem

\bibitem{lauder_sms_2019}
\begin{botherref}
\oauthor{\bsnm{{Lauder}}, \binits{G.V.}},
\oauthor{\bsnm{{Madden}}, \binits{P.G.A.}},
\oauthor{\bsnm{{Tangorra}}, \binits{J.L.}},
\oauthor{\bsnm{{Anderson}}, \binits{E.}},
\oauthor{\bsnm{{Baker}}, \binits{T.V.}}:
{Bioinspiration from fish for smart material design and function}.
Smart Material Structures
\textbf{20}(9)
(2011)
\end{botherref}
\endbibitem

\bibitem{Triantafyllou_AFM_2000}
\begin{barticle}
\bauthor{\bsnm{Triantafyllou}, \binits{M.S.}},
\bauthor{\bsnm{Triantafyllou}, \binits{G.S.}},
\bauthor{\bsnm{Yue}, \binits{D.K.P.}}:
\batitle{Hydrodynamics of fishlike swimming.}
\bjtitle{Annual Reviews of Fluid Mechanics}
\bvolume{32},
\bfpage{33}--\blpage{53}
(\byear{2000})
\end{barticle}
\endbibitem

\bibitem{Barrett1996}
\begin{botherref}
\oauthor{\bsnm{Barrett}, \binits{D.S.}}:
Propulsive efficiency of a flexible hull underwater vehicle.
PhD thesis,
Massachusetts Institute of Technology
(1996)
\end{botherref}
\endbibitem

\bibitem{pettersen_ieee_ram_2016}
\begin{barticle}
\bauthor{\bsnm{Kelasidi}, \binits{E.}},
\bauthor{\bsnm{Liljeback}, \binits{P.}},
\bauthor{\bsnm{Pettersen}, \binits{K.Y.}},
\bauthor{\bsnm{Gravdahl}, \binits{J.T.}}:
\batitle{Innovation in underwater robots: Biologically inspired swimming snake
  robots}.
\bjtitle{IEEE Robotics \& Automation Magazine}
\bvolume{23}(\bissue{1}),
\bfpage{44}--\blpage{62}
(\byear{2016})
\end{barticle}
\endbibitem

\bibitem{boyer_tor_2008}
\begin{barticle}
\bauthor{\bsnm{Boyer}, \binits{F.}},
\bauthor{\bsnm{Porez}, \binits{M.}},
\bauthor{\bsnm{Leroyer}, \binits{A.}},
\bauthor{\bsnm{Visonneau}, \binits{M.}}:
\batitle{Fast dynamics of an eel-like robot—comparisons with navier–stokes
  simulations}.
\bjtitle{IEEE Transactions on Robotics}
\bvolume{24}(\bissue{6}),
\bfpage{1274}--\blpage{1288}
(\byear{2008})
\end{barticle}
\endbibitem

\bibitem{zhu2019tuna}
\begin{barticle}
\bauthor{\bsnm{Zhu}, \binits{J.}},
\bauthor{\bsnm{White}, \binits{C.}},
\bauthor{\bsnm{Wainwright}, \binits{D.K.}},
\bauthor{\bsnm{Di~Santo}, \binits{V.}},
\bauthor{\bsnm{Lauder}, \binits{G.V.}},
\bauthor{\bsnm{Bart-Smith}, \binits{H.}}:
\batitle{Tuna robotics: A high-frequency experimental platform exploring the
  performance space of swimming fishes}.
\bjtitle{Science Robotics}
\bvolume{4}(\bissue{34}),
\bfpage{4615}
(\byear{2019})
\end{barticle}
\endbibitem

\bibitem{quinn_science_2021}
\begin{botherref}
\oauthor{\bsnm{Zhong}, \binits{Q.}},
\oauthor{\bsnm{Zhu}, \binits{J.}},
\oauthor{\bsnm{Fish}, \binits{F.E.}},
\oauthor{\bsnm{Kerr}, \binits{S.J.}},
\oauthor{\bsnm{Downs}, \binits{A.M.}},
\oauthor{\bsnm{Bart-Smith}, \binits{H.}},
\oauthor{\bsnm{Quinn}, \binits{D.B.}}:
Tunable stiffness enables fast and efficient swimming in fish-like robots.
Science Robotics
\textbf{6}(57)
(2021)
\end{botherref}
\endbibitem

\bibitem{tallapragada_ACC2015}
\begin{botherref}
\oauthor{\bsnm{Tallapragada}, \binits{P.}}:
A swimming robot with an internal rotor as a nonholonomic system.
Proceedings of the American Control Conference, 2015
(2015)
\end{botherref}
\endbibitem

\bibitem{tallapragada_kelly_jcnd2017}
\begin{barticle}
\bauthor{\bsnm{Tallapragada}, \binits{P.}},
\bauthor{\bsnm{Kelly}, \binits{S.D.}}:
\batitle{Integrability of velocity constraints modeling vortex shedding in
  ideal fluids}.
\bjtitle{Journal of Computational and Nonlinear Dynamics}
\bvolume{12}(\bissue{2}),
\bfpage{021008}
(\byear{2016})
\end{barticle}
\endbibitem

\bibitem{pt_tmech_2016}
\begin{barticle}
\bauthor{\bsnm{Pollard}, \binits{B.}},
\bauthor{\bsnm{Tallapragada}, \binits{P.}}:
\batitle{An aquatic robot propelled by an internal rotor}.
\bjtitle{IEEE/ASME Transaction on Mechatronics}
\bvolume{22}(\bissue{2}),
\bfpage{931}--\blpage{939}
(\byear{2017})
\end{barticle}
\endbibitem

\bibitem{ft_nody_2018}
\begin{botherref}
\oauthor{\bsnm{Fedonyuk}, \binits{V.}},
\oauthor{\bsnm{Tallapragada}, \binits{P.}}:
Sinusoidal control and limit cycle analysis of the dissipative chaplygin
  sleigh.
Nonlinear Dynamics
(2018)
\end{botherref}
\endbibitem

\bibitem{pft_dscc_2018}
\begin{bchapter}
\bauthor{\bsnm{Pollard}, \binits{B.}},
\bauthor{\bsnm{Fedonyuk}, \binits{V.}},
\bauthor{\bsnm{Tallapragada}, \binits{P.}}:
\bctitle{Limit Cycle Behavior and Model Reduction of an Oscillating Fish-like
  Robot}.
In: \bbtitle{Proceedings of the ASME Dynamic Systems and Control Conference}
(\byear{2018})
\end{bchapter}
\endbibitem

\bibitem{pft_nody_2019}
\begin{barticle}
\bauthor{\bsnm{Pollard}, \binits{B.}},
\bauthor{\bsnm{Fedonyuk}, \binits{V.}},
\bauthor{\bsnm{Tallapragada}, \binits{P.}}:
\batitle{Swimming on limit cycles with nonholonomic constraints}.
\bjtitle{Nonlinear Dynamics}
\bvolume{97}(\bissue{4}),
\bfpage{2453}--\blpage{2468}
(\byear{2019})
\end{barticle}
\endbibitem

\bibitem{paley_bb_2020}
\begin{barticle}
\bauthor{\bsnm{Free}, \binits{B.A.}},
\bauthor{\bsnm{Lee}, \binits{J.}},
\bauthor{\bsnm{Paley}, \binits{D.A.}}:
\batitle{Bioinspired pursuit with a swimming robot using feedback control of an
  internal rotor}.
\bjtitle{Bioinspiration and Biomimetics}
\bvolume{15}(\bissue{3}),
\bfpage{035005}
(\byear{2020})
\end{barticle}
\endbibitem

\bibitem{ft_acc_2020}
\begin{bchapter}
\bauthor{\bsnm{Fedonyuk}, \binits{V.}},
\bauthor{\bsnm{Tallapragada}, \binits{P.}}:
\bctitle{Path Tracking for the Dissipative Chaplygin Sleigh}.
In: \bbtitle{Proceedings of the American Control Conference},
pp. \bfpage{5256}--\blpage{5261}
(\byear{2020})
\end{bchapter}
\endbibitem

\bibitem{paley_acc_2020}
\begin{bchapter}
\bauthor{\bsnm{Ghanem}, \binits{P.}},
\bauthor{\bsnm{Wolek}, \binits{A.}},
\bauthor{\bsnm{Paley}, \binits{D.A.}}:
\bctitle{Planar Formation Control of a School of Robotic Fish}.
In: \bbtitle{American Control Conference}
(\byear{2020})
\end{bchapter}
\endbibitem

\bibitem{bandyopadhyay_2002}
\begin{barticle}
\bauthor{\bsnm{Bandyopadhyay}, \binits{P.R.}}:
\batitle{{Maneuvering hydrodynamics of fish and small underwater vehicles}}.
\bjtitle{Integrative and Comparative Biology}
\bvolume{42}(\bissue{1}),
\bfpage{102}--\blpage{117}
(\byear{2002})
\end{barticle}
\endbibitem

\bibitem{webb_weihs_icb_2015}
\begin{barticle}
\bauthor{\bsnm{Webb}, \binits{P.W.}},
\bauthor{\bsnm{Weihs}, \binits{D.}}:
\batitle{{Stability versus Maneuvering: Challenges for Stability during
  Swimming by Fishes}}.
\bjtitle{Integrative and Comparative Biology}
\bvolume{55}(\bissue{4}),
\bfpage{753}--\blpage{764}
(\byear{2015})
\end{barticle}
\endbibitem

\bibitem{webb_weihs_cjz_1994}
\begin{barticle}
\bauthor{\bsnm{Webb}, \binits{P.W.}},
\bauthor{\bsnm{Weihs}, \binits{D.}}:
\batitle{{Hydrostatic stability of fish with swim bladders: not all fish are
  unstable}}.
\bjtitle{Canadian Journal of Zoology}
\bvolume{72}(\bissue{6}),
\bfpage{1149}--\blpage{1154}
(\byear{1994})
\end{barticle}
\endbibitem

\bibitem{Colgate2004}
\begin{barticle}
\bauthor{\bsnm{Colgate}, \binits{J.E.}},
\bauthor{\bsnm{Lynch}, \binits{K.M.}}:
\batitle{Mechanics and control of swimming: a review}.
\bjtitle{IEEE Journal of Oceanic Engineering}
\bvolume{29},
\bfpage{660}--\blpage{673}
(\byear{2004})
\end{barticle}
\endbibitem

\bibitem{paulling_rosenberg_1959}
\begin{barticle}
\bauthor{\bsnm{Paulling}, \binits{J.R.}},
\bauthor{\bsnm{Rosenberg}, \binits{R.M.}}:
\batitle{{On Unstable Ship Motions Resulting From Nonlinear Coupling}}.
\bjtitle{Journal of Ship Research}
\bvolume{3}(\bissue{02}),
\bfpage{36}--\blpage{46}
(\byear{1959})
\end{barticle}
\endbibitem

\bibitem{newman_1979}
\begin{barticle}
\bauthor{\bsnm{Newman}, \binits{J.N.}}:
\batitle{The theory of ship motions}.
\bjtitle{Advances in Applied Mechanics}
\bvolume{18},
\bfpage{221}--\blpage{283}
(\byear{1979})
\end{barticle}
\endbibitem

\bibitem{nayfeh_jsr_1988}
\begin{barticle}
\bauthor{\bsnm{Nayfeh}, \binits{A.H.}}:
\batitle{On the undesirable roll characteristics of ships in regular seas}.
\bjtitle{Journal of Ship Research}
\bvolume{32}(\bissue{02}),
\bfpage{92}--\blpage{100}
(\byear{1988})
\end{barticle}
\endbibitem

\bibitem{neves_oe_2006}
\begin{barticle}
\bauthor{\bsnm{Neves}, \binits{M.A.}},
\bauthor{\bsnm{Rodriguez}, \binits{C.A.}}:
\batitle{On unstable ship motions resulting from strong non-linear coupling}.
\bjtitle{Ocean Engineering}
\bvolume{33}(\bissue{14-15}),
\bfpage{1853}--\blpage{1883}
(\byear{2006})
\end{barticle}
\endbibitem

\bibitem{zbm_cdc_1999}
\begin{bchapter}
\bauthor{\bsnm{Zenkov}, \binits{D.V.}},
\bauthor{\bsnm{Bloch}, \binits{A.M.}},
\bauthor{\bsnm{Marsden}, \binits{J.E.}}:
\bctitle{Stabilization of the unicycle with rider}.
In: \bbtitle{Proceedings of the 38th IEEE Conference on Decision and Control},
vol. \bseriesno{4},
pp. \bfpage{3470}--\blpage{3471}
(\byear{1999})
\end{bchapter}
\endbibitem

\bibitem{naveh_dc_1999}
\begin{barticle}
\bauthor{\bsnm{Naveh}, \binits{Y.}},
\bauthor{\bsnm{Bar-Yoseph}, \binits{P.Z.}},
\bauthor{\bsnm{Halevi}, \binits{Y.}}:
\batitle{Nonlinear modeling and control of a unicycle}.
\bjtitle{Dynamics and Control}
\bvolume{9}(\bissue{4}),
\bfpage{279}--\blpage{296}
(\byear{1999})
\end{barticle}
\endbibitem

\bibitem{deluca_2000}
\begin{barticle}
\bauthor{\bsnm{{De Luca}}, \binits{A.}},
\bauthor{\bsnm{Oriolo}, \binits{G.}},
\bauthor{\bsnm{Vendittelli}, \binits{M.}}:
\batitle{Stabilization of the unicycle via dynamic feedback linearization}.
\bjtitle{IFAC Proceedings Volumes}
\bvolume{33}(\bissue{27}),
\bfpage{687}--\blpage{692}
(\byear{2000})
\end{barticle}
\endbibitem

\bibitem{zbm_scl_2002}
\begin{barticle}
\bauthor{\bsnm{Zenkov}, \binits{D.V.}},
\bauthor{\bsnm{Bloch}, \binits{A.M.}},
\bauthor{\bsnm{Marsden}, \binits{J.E.}}:
\batitle{The lyapunov--malkin theorem and stabilization of the unicycle with
  rider}.
\bjtitle{Systems \& control letters}
\bvolume{45}(\bissue{4}),
\bfpage{293}--\blpage{302}
(\byear{2002})
\end{barticle}
\endbibitem

\bibitem{Lamb}
\begin{botherref}
\oauthor{\bsnm{Lamb}, \binits{S.H.}}:
Hydrodynamics.
Dover
(1945)
\end{botherref}
\endbibitem

\bibitem{milne-thomson96}
\begin{botherref}
\oauthor{\bsnm{Milne-Thomson}, \binits{L.M.}}:
{Theoretical Hydrodynamics}.
Dover
(1996)
\end{botherref}
\endbibitem

\bibitem{borisov_nody_2019}
\begin{barticle}
\bauthor{\bsnm{Bizyaev}, \binits{I.A.}},
\bauthor{\bsnm{Borisov}, \binits{A.V.}},
\bauthor{\bsnm{Mamaev}, \binits{I.S.}}:
\batitle{Dynamics of a chaplygin sleigh with an unbalanced rotor: regular and
  chaotic motions}.
\bjtitle{Nonlinear Dynamics}
\bvolume{98}(\bissue{3}),
\bfpage{2277}--\blpage{2291}
(\byear{2019})
\end{barticle}
\endbibitem

\bibitem{Mathieu_RRand}
\begin{botherref}
\oauthor{\bsnm{Kovacic}, \binits{I.}},
\oauthor{\bsnm{Richard}, \binits{R.}},
\oauthor{\bsnm{Sah}, \binits{M.S.}}:
Mathieu's equation and its generalizations: Overview of stability charts and
  their features.
Applied Mechanics Reviews
\textbf{70}(2)
(2018).
\doiurl{10.1115/1.4039144}.
020802
\end{botherref}
\endbibitem

\bibitem{nayfeh2008perturbation}
\begin{botherref}
\oauthor{\bsnm{Nayfeh}, \binits{A.H.}}:
Perturbation methods.
John Wiley and Sons, Ltd
(2000)
\end{botherref}
\endbibitem

\bibitem{yakubovich_1975}
\begin{bbook}
\bauthor{\bsnm{Yakubovich}, \binits{V.}},
\bauthor{\bsnm{Starzhinskii}, \binits{V.}}:
\bbtitle{Linear Differential Equations with Periodic Coefficients}.
\bpublisher{Wiley},
\blocation{New York}
(\byear{1975})
\end{bbook}
\endbibitem

\bibitem{Stroker_nonlinearvibr}
\begin{bbook}
\bauthor{\bsnm{Stoker}, \binits{J.J.}}:
\bbtitle{Nonlinear Vibrations in Mechanical and Electrical}.
\bpublisher{Interscience},
\blocation{Publishers}
(\byear{1950})
\end{bbook}
\endbibitem

\bibitem{Magnus_Winkler_Hillseqn}
\begin{botherref}
\oauthor{\bsnm{Magnus}, \binits{W.}},
\oauthor{\bsnm{Stanley}, \binits{W.}}:
Hill's equation.
Interscience Publishers
(1966)
\end{botherref}
\endbibitem

\bibitem{Burgh2002AnEW}
\begin{botherref}
\oauthor{\bparticle{van~der} \bsnm{Burgh}, \binits{A.H.P.}}:
An equation with a time-periodic damping coefficient: Stability diagram and an
  application.
Reports of the Department of Applied Mathematical Analysis
(2002)
\end{botherref}
\endbibitem

\bibitem{Afzali2017AnalysisOT}
\begin{bchapter}
\bauthor{\bsnm{Afzali}, \binits{F.}},
\bauthor{\bsnm{Acar}, \binits{G.}},
\bauthor{\bsnm{Feeny}, \binits{B.F.}}:
\bctitle{Analysis of the periodic damping coefficient equation based on floquet
  theory}.
(\byear{2017})
\end{bchapter}
\endbibitem

\bibitem{Batchelor_damp}
\begin{barticle}
\bauthor{\bsnm{Batchelor}, \binits{D.B.}}:
\batitle{Parametric resonance of systems with time‐varying dissipation}.
\bjtitle{Applied Physics Letters}
\bvolume{29}(\bissue{5}),
\bfpage{280}--\blpage{281}
(\byear{1976})
\end{barticle}
\endbibitem

\bibitem{YOUNESIAN200758}
\begin{barticle}
\bauthor{\bsnm{Younesian}, \binits{D.}},
\bauthor{\bsnm{Esmailzadeh}, \binits{E.}},
\bauthor{\bsnm{Sedaghati}, \binits{R.}}:
\batitle{Asymptotic solutions and stability analysis for generalized
  non-homogeneous mathieu equation}.
\bjtitle{Communications in Nonlinear Science and Numerical Simulation}
\bvolume{12}(\bissue{1}),
\bfpage{58}--\blpage{71}
(\byear{2007})
\end{barticle}
\endbibitem

\bibitem{SHADMAN200568}
\begin{barticle}
\bauthor{\bsnm{Shadman}, \binits{D.}},
\bauthor{\bsnm{Mehri}, \binits{B.}}:
\batitle{A non-homogeneous hill’s equation}.
\bjtitle{Applied Mathematics and Computation}
\bvolume{167}(\bissue{1}),
\bfpage{68}--\blpage{75}
(\byear{2005})
\end{barticle}
\endbibitem

\bibitem{Slane2011AnalysisOP}
\begin{barticle}
\bauthor{\bsnm{Slane}, \binits{J.H.}},
\bauthor{\bsnm{Tragesser}, \binits{S.G.}}:
\batitle{Analysis of periodic nonautonomous i nhomogeneous systems}.
\bjtitle{Nonlinear dynamics and systems theory}
\bvolume{11},
\bfpage{183}--\blpage{198}
(\byear{2011})
\end{barticle}
\endbibitem

\bibitem{Rodriguez2015OnSO}
\begin{botherref}
\oauthor{\bsnm{Rodriguez}, \binits{A.}},
\oauthor{\bsnm{Collado}, \binits{J.}}:
On stability of periodic solutions in non-homogeneous hill's equation.
2015 12th International Conference on Electrical Engineering, Computing Science
  and Automatic Control (CCE),
1--6
(2015)
\end{botherref}
\endbibitem

\bibitem{Webb2001}
\begin{barticle}
\bauthor{\bsnm{Webb}, \binits{D.C.}},
\bauthor{\bsnm{Simonetti}, \binits{P.J.}},
\bauthor{\bsnm{Jones}, \binits{C.P.}}:
\batitle{Slocum: An underwater glider propelled by environmental energy}.
\bjtitle{IEEE Journal of Oceanic Engineering}
\bvolume{26},
\bfpage{447}--\blpage{452}
(\byear{2001})
\end{barticle}
\endbibitem

\end{thebibliography}


\end{document}